# Shotgun DNA sequencing evidence: sample-specific and unknown genotyping error probabilities

Mikkel Meyer Andersen[a,b,*]

[a] *Department of Mathematical Sciences, Aalborg University, Denmark,*

[b] *Section of Forensic Genetics, Department of Forensic Medicine, Faculty of Health and Medical Sciences, University of Copenhagen, Denmark,*

## Abstract

Many forensic genetic trace samples are of too low quality to obtain short tandem repeat (STR) DNA profiles as the nuclear DNA they contain is highly degraded (e.g., telogen hairs). Instead, performing shotgun DNA sequencing of such samples can provide valuable information on, e.g., single nucleotide polymorphism (SNP) markers. As a result, shotgun sequencing is starting to gain more attention in forensic genetics and statistical models to correctly interpret such evidence, including properly accounting for sequencing errors, are needed. One such model is the `wgsLR` model by Andersen et. al. (2025) that enabled evaluating the evidential strength of a comparison between the genotypes in the trace sample and reference sample assuming a single-source contribution to both samples. This paper extends the `wgsLR` model to allow for different (asymmetric) genotyping error probabilities (e.g., from a low quality trace sample and a high quality reference sample). The model was also extended to handle unknown genotyping error probabilities via both maximising profile likelihood and using a prior distribution. The sensitivity of the `wgsLR` model against overdispersion was also investigated and it was found robust against it. It was also found that handling an unknown genotyping error probability of the trace sample with the methods having a sufficient number of independent markers gave concordant weight of evidence (WoE) under both the hypotheses (same or different individuals being donors of trace and reference sample). It was found more conservative to use a too small trace sample genotyping error probability rather than a too high genotyping error probability as the latter can explain genotype inconsistencies by errors rather than due to two different individuals being the donors of the trace sample and reference sample. The extensions of the model are implemented in the `R` package `wgsLR`.

## 1. Introduction

Forensic genetics aims to be able to analyse and interpret all possible biological trace samples for genetic evidence. Starting with analysing and interpreting single-source trace samples years ago, it is now possible, with vast improvements at multiple fronts (e.g., procedures, apparatus, and statistical methods for interpretation), to analyse mixtures from multiple contributors, even from small amounts of of DNA [1].

Most standard forensic genetic DNA analyses are performed to obtain short tandem repeat (STR) DNA profiles using polymerase chain reaction capillary electrophoresis (PCR-CE) with amplicons of size 85-400 bp [1, 2, 3].

Still, many trace samples are of such low DNA quantity or quality (e.g., very fragmented DNA) that it is not possible to obtain STR DNA profiles using PCR-CE. A recent study states that 50% of the samples received in Denmark between 2016 and 2022 were deemed unsuitable for standard PCR-CE analyses due to insufficient DNA quantity or quality [4]. Telogen hairs is an examples of such trace samples where it is not possible to obtain STR DNA profiles by PCR-CE [5].

*Corresponding author
*Email address:* `mikl@math.aau.dk` (Mikkel Meyer Andersen)

Instead, performing shotgun DNA sequencing of such trace samples (of too low quality to obtain STR DNA profiles by PCR-CE) is starting to gain more attention in forensic genetics as the DNA fragments that are there, typically of size less than 100 bp, can still provide valuable information on, e.g., single nucleotide polymorphism (SNP) markers [5, 6].

As shotgun DNA sequencing is not error-free [5], the situation is that statistical interpretation models taking asymmetric sequencing error probabilities for the trace sample and reference sample are needed. Different statistical models have been proposed, both some that models the raw sequencing data [7, 8, 9] as well as the data with called genotypes [5] that can also be used to estimate the genotyping error probability.

In this paper the `wgsLR` model by [5] to evaluate the evidential value of single-source samples is extended. To explain in what ways, first note that a central parameter of the model by [5] was the genotyping error probability that was denoted by $w$. The original model [5] was used both to estimate $w$ using duplicated samples from the same individual and to calculate the weight of evidence in terms of a likelihood ratio ($LR$) assuming the same genotyping error probability, $w$, for both the trace sample and reference sample.

This paper contains three main results. Firstly, it is investigated if the `wgsLR` model is robust against overdispersion. That is, how well can $w$ be estimated as long as the average genotyping error probability across the genome is $w$ even though some genomic ranges have higher genotyping error probability and other genomic ranges lower genotyping error probability - as long as the genotyping error probability on average is $w$? This is important for when do estimate the genotyping error probability parameter.

Secondly, the model is extended to handle sample-specific genotyping error probabilities. This is important as the **t**race sample, $X_t$, can be of poor quality, and thereby have a higher value of $w$ than a good quality **r**eference sample, $X_r$, from a person of interest (PoI), e.g., a suspect. The reference sample can typically, in contrast to the trace sample, be analysed to any desired quality as there are plenty of high quality DNA available. The genotyping error probability for the trace sample is referred to as $w_t$ and the genotyping error probability for the reference sample is referred to as $w_r$. This extension enables $LR$ computations for sample-specific genotyping error probabilities. In this paper the consequences of such asymmetric genotyping error probabilities.

Thirdly, different approaches to calculate an $LR$ even when no good estimates of $w$ are available are investigated. This is done by using a weighted average over a prior distribution of potential known trace sample genotyping error probabilities (in Bayesian context this is called the prior predictive distribution) and by maximising the profile likelihood under each hypothesis (finding the value of the trace sample genotyping error probability that maximise each of the likelihoods).

## 2. Method

All simulations and data analysis was performed in `R` [10] version 4.3.3 using `tidyverse` [11].

As in the original description of the model [5], only binary single nucleotide polymorphisms (SNPs) are considered and are coded as 0, 1, and 2 minor alleles with genotype probabilities $p_0$, $p_1$, and $p_2$, respectively. The true, unobserved genotype from the trace sample donor was originally in [5] denoted by $Z^D \in \{0, 1, 2\}$ (D for donor) and the observed genotype (potentially with errors) was denoted by $X^D$. Similarly, the unobserved genotype from the person of interest was denoted by $Z^S \in \{0, 1, 2\}$ (S for suspect) and the observed genotype (potentially with errors) was denoted by $X^S$. Here, the notation is updated to $t$ (instead of $D$) to refer to the **t**race sample and $r$ for **r**eference sample. Thus, the genotyping error probability for the trace sample is denoted $w_t$ and the genotyping error probability for the reference sample is denoted $w_r$. The reason of the update of the notation is that there has been some confusion about whether the donor sample was referring to the trace or reference sample. The hope is that the updated notation with trace and reference is more clear for everybody as it refers to the samples. The `R` package `wgsLR` has also been refactored to reflect this update, and the development version is available at https://github.com/mikldk/wgsLR.

In the following, the three aspects studied (overdispersion, sample-specific genotyping error probabilities, and unknown genotyping error probabilities) will be described followed by a section of describing the simulations performed.

### *2.1. Overdispersion*

If each nucleotide in the nuclear genome has a genotyping error probability of $w$, then $w$ can be estimated using the `wgsLR` model as shown by [5]. This was shown for fixed $w$ for all nucleotides. If some genomic regions are more likely to have a genotyping error than other genomic regions and the mean of the genotyping error probability is still $w$, this is called overdispersion (more variance than the model expects). Here, the sensitivity of overdispersion of the estimator in [5] is investigated. For each position, a genotyping error probability was drawn randomly from a beta distribution (on $(0, \frac{1}{2})$) with mean value $w$ for different choices of variances, $\sigma^2$. The different beta distributions (on $(0, \frac{1}{2})$) used are displayed in Figure 1. Only one value of mean value of $w$ was used. A lower mean value would increase the requirement of the sample size to obtain the same precision as larger mean values.

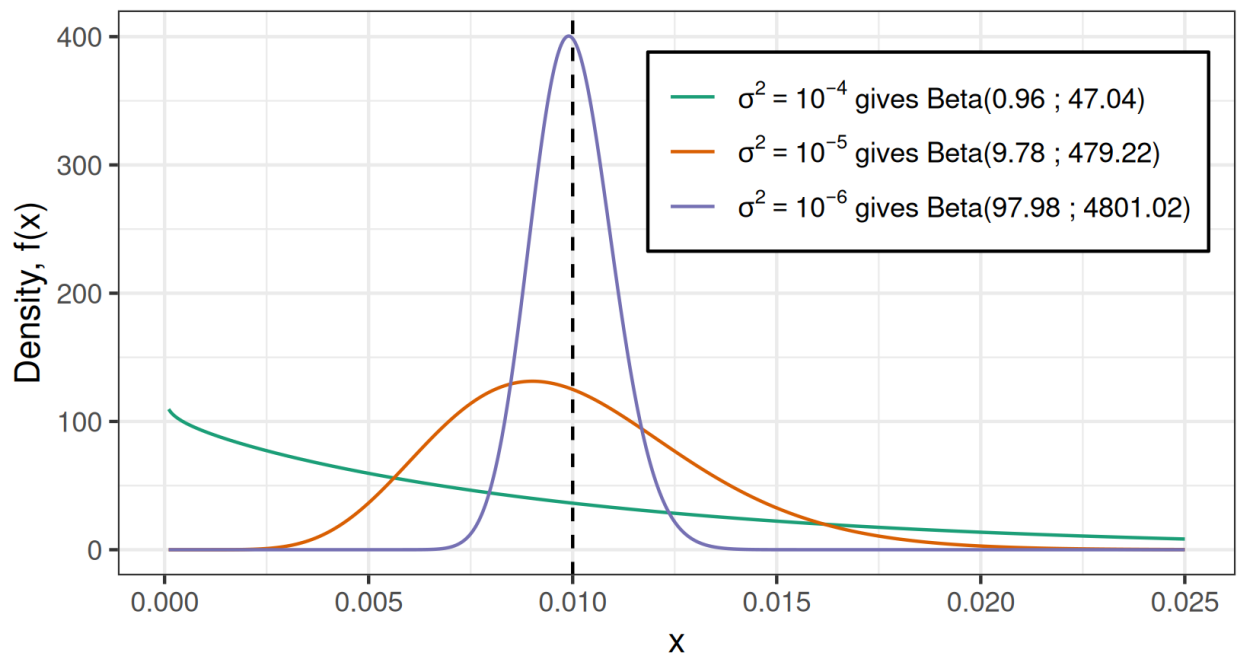

Figure 1: Beta distributions used in study of effect of overdispersion. All the beta distributions have support on $(0, \frac{1}{2})$. Vertical dashed line shows mean value 0.01 (same for all three distributions). The variance, denoted by $\sigma^2$, is different for each distribution. The $\alpha$ and $\beta$ numbers in Beta($\alpha$; $\beta$) denotes the shape parameters for the beta distributions obtained from the mean and variance for the distribution.

### *2.2. Sample-specific genotyping error probabilities*

To extend the `wgsLR` model by [5] to handle sample-specific genotyping error probabilities, $w_t$ and $w_r$, the graphical representation of the model can be revisited, and for each $w$, it can be assigned if an error $(w_t)$/non-error $(1 - w_t)$ in the trace sample is needed or if it is related to the reference sample ($w_r$ and $1 - w_r$ for error and non-error, respectively). This can be done systematically, and was done in the `wgsLR` `R` package (available at https://github.com/mikldk/wgsLR) to obtain the formulas for the $LR$ with sample-specific genotyping error probabilities, $w_t$ and $w_r$, that can be found in Table 1. The formulas can be accessed via the `R` package `wgsLR` via the objects `d_LR_w` for the same $w$ for both samples and `d_LR_wTwR` for sample-specific $w_t$ and $w_r$ for trace and reference sample, respectively.

Table 1: $LR$ for sample-specific genotyping error probabilities, $w_t$ (for trace sample) and $w_r$ (for reference sample). The formulas can be found in the object `d_LR_wTwR` in the `wgsLR` R package [5]. The formulas for using the same genotyping error probability for both samples as originally introduced in [5] can be found in the object `d_LR_w` (previously the object `d_prob_LR`). Note, the postfix indicate if there is one common or two, sample-specific, genotyping error probabilities.

| $\mathbf{X_t}$ | $\mathbf{X_r}$ | **LR** |
|---|---|---|
| 0 | 0 | $\frac{p_0(w_r-1)^2(w_t-1)^2+p_1 w_r w_t(w_r-1)(w_t-1)+p_2 w_r^2 w_t^2}{p_0^2(w_r-1)^2(w_t-1)^2-p_0p_1 w_r(w_r-1)(w_t-1)^2-p_0p_1 w_t(w_r-1)^2(w_t-1)+p_0p_2 w_r^2(w_t-1)^2+p_0p_2 w_t^2(w_r-1)^2+p_1^2 w_r w_t(w_r-1)(w_t-1)-p_1p_2 w_r^2 w_t(w_t-1)-p_1p_2 w_r w_t^2(w_r-1)+p_2^2 w_r^2 w_t^2}$ |
| 0 | 1 | $\frac{2p_0 w_r(w_r-1)(w_t-1)^2+p_1 w_r^2 w_t(w_t-1)+p_1 w_t(w_r-1)^2(w_t-1)+2p_2 w_r w_t^2(w_r-1)}{2p_0^2 w_r(w_r-1)(w_t-1)^2-p_0p_1 w_r^2(w_t-1)^2-2p_0p_1 w_r w_t(w_r-1)(w_t-1)-p_0p_1(w_r-1)^2(w_t-1)^2+2p_0p_2 w_r w_t^2(w_r-1)+2p_0p_2 w_r(w_r-1)(w_t-1)^2+p_1^2 w_r^2 w_t(w_t-1)+p_1^2 w_t(w_r-1)^2(w_t-1)-p_1p_2 w_r^2 w_t^2-2p_1p_2 w_r w_t(w_r-1)(w_t-1)-p_1p_2 w_t^2(w_r-1)^2+2p_2^2 w_r w_t^2(w_r-1)}$ |
| 0 | 2 | $\frac{p_0 w_r^2(w_t-1)^2+p_1 w_r w_t(w_r-1)(w_t-1)+p_2 w_t^2(w_r-1)^2}{p_0^2 w_r^2(w_t-1)^2-p_0p_1 w_r^2 w_t(w_t-1)-p_0p_1 w_r(w_r-1)(w_t-1)^2+p_0p_2 w_r^2 w_t^2+p_0p_2(w_r-1)^2(w_t-1)^2+p_1^2 w_r w_t(w_r-1)(w_t-1)-p_1p_2 w_r w_t^2(w_r-1)-p_1p_2 w_t(w_r-1)^2(w_t-1)+p_2^2 w_t^2(w_r-1)^2}$ |
| 1 | 0 | $\frac{2p_0 w_t(w_r-1)^2(w_t-1)+p_1 w_r w_t^2(w_r-1)+p_1 w_r(w_r-1)(w_t-1)^2+2p_2 w_r^2 w_t(w_t-1)}{2p_0^2 w_t(w_r-1)^2(w_t-1)-2p_0p_1 w_r w_t(w_r-1)(w_t-1)-p_0p_1 w_t^2(w_r-1)^2-p_0p_1(w_r-1)^2(w_t-1)^2+2p_0p_2 w_r^2 w_t(w_t-1)+2p_0p_2 w_t(w_r-1)^2(w_t-1)+p_1^2 w_r w_t^2(w_r-1)+p_1^2 w_r(w_r-1)(w_t-1)^2-p_1p_2 w_r^2 w_t^2-p_1p_2 w_r^2(w_t-1)^2-2p_1p_2 w_r w_t(w_r-1)(w_t-1)+2p_2^2 w_r^2 w_t(w_t-1)}$ |
| 1 | 1 | $\frac{-4p_0 w_r w_t(w_r-1)(w_t-1)-p_1 w_r^2 w_t^2-p_1 w_r^2(w_t-1)^2-p_1 w_t^2(w_r-1)^2-p_1(w_r-1)^2(w_t-1)^2-4p_2 w_r w_t(w_r-1)(w_t-1)}{-4p_0^2 w_r w_t(w_r-1)(w_t-1)+2p_0p_1 w_r^2 w_t(w_t-1)+2p_0p_1 w_r w_t^2(w_r-1)+2p_0p_1 w_r(w_r-1)(w_t-1)^2+2p_0p_1 w_t(w_r-1)^2(w_t-1)-8p_0p_2 w_r w_t(w_r-1)(w_t-1)-p_1^2 w_r^2 w_t^2-p_1^2 w_r^2(w_t-1)^2-p_1^2 w_t^2(w_r-1)^2-p_1^2(w_r-1)^2(w_t-1)^2+2p_1p_2 w_r^2 w_t(w_t-1)+2p_1p_2 w_r w_t^2(w_r-1)+2p_1p_2 w_r(w_r-1)(w_t-1)^2+2p_1p_2 w_t(w_r-1)^2(w_t-1)-4p_2^2 w_r w_t(w_r-1)(w_t-1)}$ |
| 1 | 2 | $\frac{2p_0 w_r^2 w_t(w_t-1)+p_1 w_r w_t^2(w_r-1)+p_1 w_r(w_r-1)(w_t-1)^2+2p_2 w_t(w_r-1)^2(w_t-1)}{2p_0^2 w_r^2 w_t(w_t-1)-p_0p_1 w_r^2 w_t^2-p_0p_1 w_r^2(w_t-1)^2-2p_0p_1 w_r w_t(w_r-1)(w_t-1)+2p_0p_2 w_r^2 w_t(w_t-1)+2p_0p_2 w_t(w_r-1)^2(w_t-1)+p_1^2 w_r w_t^2(w_r-1)+p_1^2 w_r(w_r-1)(w_t-1)^2-2p_1p_2 w_r w_t(w_r-1)(w_t-1)-p_1p_2 w_t^2(w_r-1)^2-p_1p_2(w_r-1)^2(w_t-1)^2+2p_2^2 w_t(w_r-1)^2(w_t-1)}$ |
| 2 | 0 | $\frac{p_0 w_t^2(w_r-1)^2+p_1 w_r w_t(w_r-1)(w_t-1)+p_2 w_r^2(w_t-1)^2}{p_0^2 w_t^2(w_r-1)^2-p_0p_1 w_r w_t^2(w_r-1)-p_0p_1 w_t(w_r-1)^2(w_t-1)+p_0p_2 w_r^2 w_t^2+p_0p_2(w_r-1)^2(w_t-1)^2+p_1^2 w_r w_t(w_r-1)(w_t-1)-p_1p_2 w_r^2 w_t(w_t-1)-p_1p_2 w_r(w_r-1)(w_t-1)^2+p_2^2 w_r^2(w_t-1)^2}$ |
| 2 | 1 | $\frac{2p_0 w_r w_t^2(w_r-1)+p_1 w_r^2 w_t(w_t-1)+p_1 w_t(w_r-1)^2(w_t-1)+2p_2 w_r(w_r-1)(w_t-1)^2}{2p_0^2 w_r w_t^2(w_r-1)-p_0p_1 w_r^2 w_t^2-2p_0p_1 w_r w_t(w_r-1)(w_t-1)-p_0p_1 w_t^2(w_r-1)^2+2p_0p_2 w_r w_t^2(w_r-1)+2p_0p_2 w_r(w_r-1)(w_t-1)^2+p_1^2 w_r^2 w_t(w_t-1)+p_1^2 w_t(w_r-1)^2(w_t-1)-p_1p_2 w_r^2(w_t-1)^2-2p_1p_2 w_r w_t(w_r-1)(w_t-1)-p_1p_2(w_r-1)^2(w_t-1)^2+2p_2^2 w_r(w_r-1)(w_t-1)^2}$ |
| 2 | 2 | $\frac{p_0 w_r^2 w_t^2+p_1 w_r w_t(w_r-1)(w_t-1)+p_2(w_r-1)^2(w_t-1)^2}{p_0^2 w_r^2 w_t^2-p_0p_1 w_r^2 w_t(w_t-1)-p_0p_1 w_r w_t^2(w_r-1)+p_0p_2 w_r^2(w_t-1)^2+p_0p_2 w_t^2(w_r-1)^2+p_1^2 w_r w_t(w_r-1)(w_t-1)-p_1p_2 w_r(w_r-1)(w_t-1)^2-p_1p_2 w_t(w_r-1)^2(w_t-1)+p_2^2(w_r-1)^2(w_t-1)^2}$ |

### 2.3. Unknown genotyping error probabilities

Calculating the $LR$s given in Table 1 assumes that $w_t$ and $w_r$ are known. In practice this assumption is adhered to for the reference sample as that can be sufficiently determined in the lab (e.g. for buccal swabs), and thus $w_r$ can be assumed known.

#### 2.3.1. Integration over prior distribution

On the other hand, because the conditions and quality of the trace sample is unknown, it can be problematic to determine $w_t$. Thus, instead a Bayesian standpoint can be taken and view $w_t$ as a random variable that follows some (prior) probability distribution. Recall that

$$LR = \frac{P(E \mid H_1)}{P(E \mid H_2)}, \tag{1}$$

where $H_1$ is that the same individual donated both the trace sample and the reference sample and $H_2$ is that two different individuals donated the trace sample and reference sample. Then for each $H_i$ this can be obtained by marginalising out $w_t$ according to a prior distribution $P_i(w_t)$ of $w_t$, i.e.

$$P(E \mid H_i) = \int P(E \mid H_i, w_t) P_i(w_t) \, \mathrm{d}w_t = \mathbf{E}_{w_t \sim P_i} \left[ P(E \mid H_i, w_t) \right]. \tag{2}$$

This is sometimes called the prior predictive distribution of the evidence $E$ under the hypothesis $H_i$. It is also the expected value of the likelihood function $P(E \mid H_i, w_t)$, where the expectation is taken with respect to the prior distribution $P_i(w_t)$ of $w_t$. Intuitively, it is a weighted average of the likelihood, so the likelihood is calculated for each $w_t$ and is weighted by how likely $w_t$ is *a priori*.

Thus,

$$\begin{aligned} LR &= \frac{P(E \mid H_1)}{P(E \mid H_2)} && (3) \\ &= \frac{\int P(E \mid H_1, w_t) P_1(w_t) \, \mathrm{d}w_t}{\int P(E \mid H_2, w_t) P_2(w_t) \, \mathrm{d}w_t} && (4) \\ &= \frac{\mathbb{E}_{w_t \sim P_1} \left[ P(E \mid H_1, w_t) \right]}{\mathbb{E}_{w_t \sim P_2} \left[ P(E \mid H_2, w_t) \right]}. && (5) \end{aligned}$$

For multiple independent SNP markers, $E = (E_1, E_2, \ldots, E_m)$, it is possible to either integrate the product (assuming the same $w_t$ for all $m$ markers) or take a product of the integrals (allowing the markers to have their own $w_t$). It will now be shown that instead of calculating the $LR$s, then using WoE $= \log_{10}(LR)$, results in no need for assumptions (same/different $w_t$ for markers and/or the hypotheses). This is shown as follows:

$$\begin{aligned} WoE &= \mathbb{E}_{w_t \sim P} \left[ \log_{10} \left( \prod_{j=1}^{m} \frac{P(E_j \mid H_1, w_t)}{P(E_j \mid H_2, w_t)} \right) \right] && (6) \\ &= \mathbb{E}_{w_t \sim P} \left[ \sum_{j=1}^{m} \log_{10} \left( \frac{P(E_j \mid H_1, w_t)}{P(E_j \mid H_2, w_t)} \right) \right] && (7) \\ &= \mathbb{E}_{w_t \sim P} \left[ \sum_{j=1}^{m} \log_{10} P(E_j \mid H_1, w_t) - \sum_{j=1}^{m} \log_{10} P(E_j \mid H_2, w_t) \right] && (8) \\ &= \sum_{j=1}^{m} \mathbb{E}_{w_t \sim P} \left[ \log_{10} P(E_j \mid H_1, w_t) \right] - \sum_{j=1}^{m} \mathbb{E}_{w_t \sim P} \left[ \log_{10} P(E_j \mid H_2, w_t) \right] && (9) \\ &= \sum_{j=1}^{m} \int \log_{10} \left( P(E_j \mid H_1, w_t) \right) P(w_t) \, \mathrm{d}w_t - \sum_{j=1}^{m} \int \log_{10} \left( P(E_j \mid H_2, w_t) \right) P(w_t) \, \mathrm{d}w_t. && (10) \end{aligned}$$

Note that the sum of the expectations and the expectation of the sum is the same, and similarly, the sum of the integral and the integral of the sum is the same. Thus, there is only need for one prior distribution of $w_t$, and there is no need to assume the same nor different $w_t$'s across the markers as the result is the same.

In the software `wgsLR`, it is assumed *a priori* under $H_i$ that $w_t$ follows a beta distribution on $(0, \frac{1}{2})$ with shape parameters $\alpha$ and $\beta$.

Uncertainty in $w_r$ can also be handled in a similar way. This has been left out intentionally as this would not normally be the case, and it may raise problems if both genotyping error probabilities (from trace sample, $w_t$, and reference sample, $w_r$) are unknown. Also, the genotyping error probability for reference samples can be found in the lab, e.g., by making a study of multiple reference samples to determine the genotyping error probability.

*2.3.2. Maximising profile likelihood*

Another approach is to use the $w_t$ that maximise $P(E \mid H_i, w_t)$ under each hypothesis $i$. For multiple markers, it is here assumed that the same $w_t$ was used for all markers, such that

$$WoE = \log_{10}\left(\frac{\max_{w_t}\prod_{j=1}^{m} P(E_j \mid H_1, w_t)}{\max_{w_t}\prod_{j=1}^{m} P(E_j \mid H_2, w_t)}\right) \tag{11}$$

$$= \max_{w_t}\log_{10}\left(\prod_{j=1}^{m} P(E_j \mid H_1, w_t)\right) - \max_{w_t}\log_{10}\left(\prod_{j=1}^{m} P(E_j \mid H_2, w_t)\right) \tag{12}$$

$$= \max_{w_t}\sum_{j=1}^{m}\log_{10}\left(P(E_j \mid H_1, w_t)\right) - \max_{w_t}\sum_{j=1}^{m}\log_{10}\left(P(E_j \mid H_2, w_t)\right) \tag{13}$$

as the log of the maximum of a function is the same as maximising the log of the function. The maximum is taken over the interval $(w_t, \frac{1}{2})$. Note that the maximum of a sum is in general not the same as the sum of the maxima, and hence the assumption about the same $w_t$ across markers is still needed.

*2.3.3. Summary*

In summary, there are multiple ways to handle unknown $w_t$ in calculating WoE = $\log_{10}(LR)$. Here the following two methods were considered: Firstly, by assuming a prior distribution of $w_t$ and calculating a weighted average as shown in (7) and is implemented in `wgsLR` in the methods `calc_WoE_wTwR_integrate_wT_mc()` using Monte Carlo integration and `calc_WoE_wTwR_integrate_wT_num()` using numerical integration. Secondly, by using a profile likelihood in which each hypothesis can choose the $w_t$ that maximise its likelihood as shown in (11) and is implemented in `wgsLR` in the method `calc_WoE_wTwR_profilemax_wT_num()`.

To analyse the sensitivity of wrongly specified prior distribution of $w_t$, for example mean value much lower or much higher than the correct value, simulation experiments were conducted using `wgsLR` (cf. below).

In this study integration was performed with Monte Carlo integration using 1,000 samples from the prior.

*2.4. Simulations*

The simulations were performed assuming Hardy-Weinberg equilibrium with allele frequencies $q = 0.75$ (resulting in genotype frequencies $(p_0, p_1, p_2) = \left(q^2, 2q(1-q), (1-q)^2\right) = (0.5625, 0.375, 0.0625)$ for 0, 1, and 2 reference alleles, respectively) and $q = 0.9$ (resulting in $(p_0, p_1, p_2) = (0.81, 0.18, 0.01)$).

For the study of overdispersion, $w$ was simulated from the beta distributions (on $(0, \frac{1}{2})$) displayed in Figure 1. They all have a mean value of 0.01. Tables of sizes 1,000, 10,000, and 100,000 were simulated. Each combination were replicated 1,000 times. Thus, a total of 1,000 × 2 ($q$) × 3 ($w$ variance) × 3 (table sizes) = 18,000 tables were simulated.

For the study on integrating out $w_t$, the same $q$s were used. A fixed $w_r = 10^{-4}$ was used for all simulations. Different values of $w_t$ were used; namely $w_t = w_r = 10^{-4}$, $w_t = 10w_r = 10^{-3}$, and $w_t = 100w_r = 10^{-2}$ to indicate that the trace sample typically has a higher genotyping error probability than the reference sample. The number of independent SNP markers simulated were 50, 100, and 200. Each combination were replicated 1,000 times. In each combination, a “case” under $H_1$ (donor to trace sample and reference sample is the same individual) and a case under $H_2$ (donors to trace sample and reference sample are two different individuals) were simulated. Thus, a total of 1,000 $\times$ 2 ($q$) $\times$ 3 (true $w_t$ values) $\times$ 3 (number of markers) = 18,000 cases under $H_1$ were simulated and the same for $H_2$. For each case, a number of different prior distributions on $w_t$ were used in the integration, including a uniform prior on $(0, \frac{1}{2})$; a prior distribution with a mean value, $\mu$, corresponding to the true $w_t$ used in the simulations; and some with smaller and bigger mean values to investigate if it was conservative to underestimate or overestimate the genotyping error probability. The variance parameter $\sigma^2$ was chosen to be $\sigma^2 = \mu^2/2$ in one prior distribution and $\sigma^2 = 2mu^2$ in another. From the mean value and variance, the shape parameters were found. The parameters for the beta priors (on $(0, \frac{1}{2})$) are shown in Table 2 and visualised in Figure 2. Also, underestimating $w_t$ by using $w_t = w_r$ was tried, i.e., how the results were if simply assuming that $w_t = w_r$ even though the true $w_t$ was higher. In all cases, the weight of evidence (WoE) was calculated using the formulas in Table 1 via the `R` package `wgsLR`.

The different methods for handling unknown $w_t$ (the true, but unknown $w_t$ value as an oracle; using $w_t = w_r$; maximising profile likelihood; and integration over different prior distributions) were compared both by summarising how often the sign of the WoE was correct, i.e., WoE $> 0$ for $H_1$ cases (donor to trace sample and reference sample is the same individual) and WoE $< 0$ for $H_2$ cases (donors to trace sample and reference sample are two different individuals) and by calculating the empirical cross-entropy (ECE) assuming a prior odds of 1.

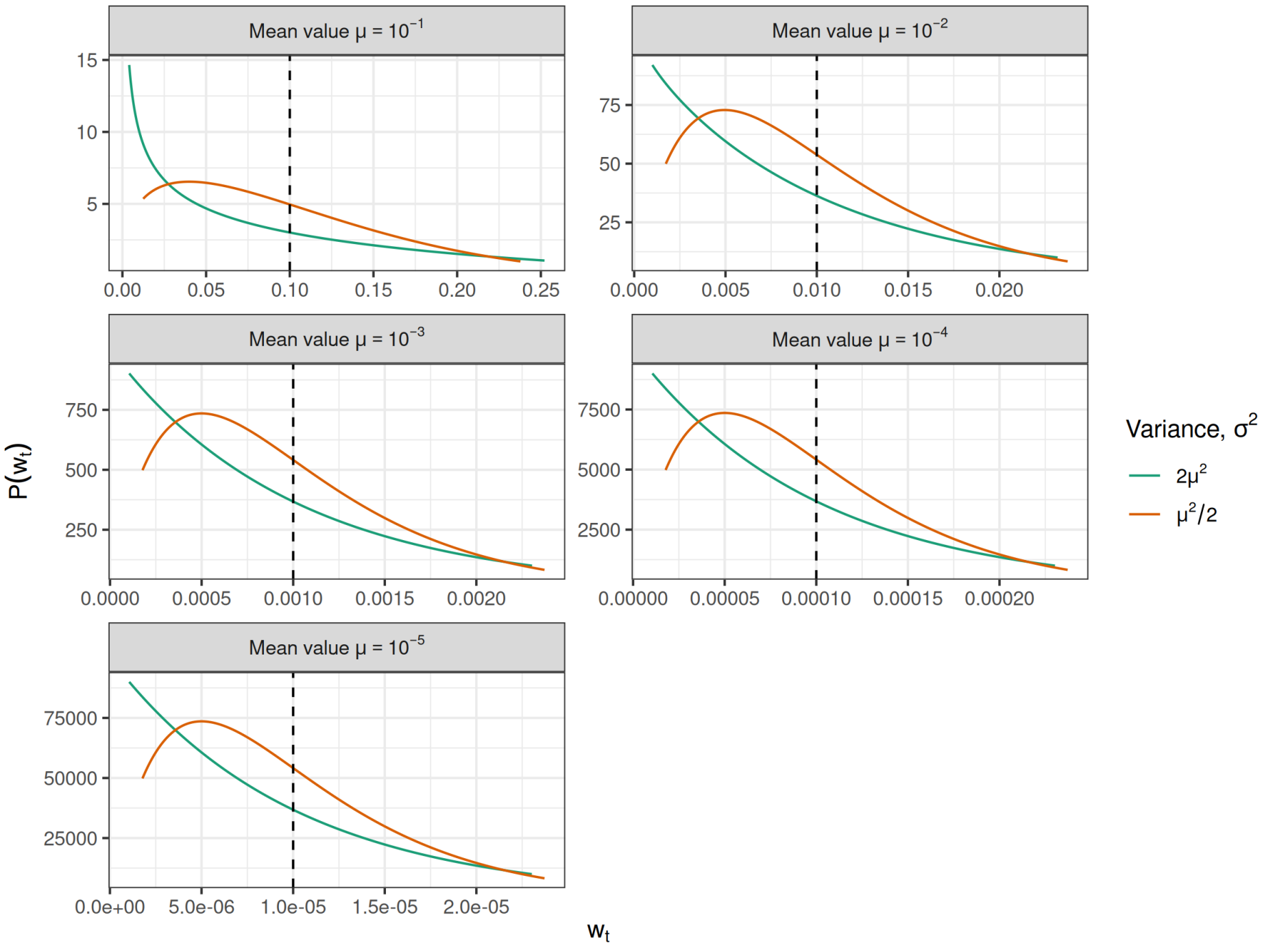

Figure 2: Prior distributions used for the analysis of integrating out $w_t$. All are beta distributions on $(0, \frac{1}{2})$. The shape parameters are calculated from the given values for the mean and variance. See more details (like numeric value for variances, shape parameters, and fractiles) in Table 2.

## 3. Results

### 3.1. Overdispersion

The sensitivity of the `wgsLR` model [5] against overdispersion was found to be very robust against overdispersion in estimating $w$, i.e., the correct mean value of $w$ was recovered even with overdispersion (Figure 3).

### 3.2. Unknown genotyping error probabilities

Table 2: Prior distributions used for the analysis of integrating out $w_t$. All are beta distributions on $(0, \frac{1}{2})$. The shape parameters are calculated from the given values for the mean and variance.

| Mean | Type | Variance | Shape 1 | Shape 2 | 5% fractile | 95% fractile |
|---|---|---|---|---|---|---|
| $10^{-1}$ | $\mu^2/2$ | $5 \times 10^{-3}$ | 1.400 | 5.600 | 0.0124384 | 0.2377969 |
| $10^{-1}$ | $2\mu^2$ | $10^{-2}$ | 0.600 | 2.400 | 0.0012744 | 0.3103297 |
| $10^{-2}$ | $\mu^2/2$ | $5 \times 10^{-5}$ | 1.940 | 95.060 | 0.0017324 | 0.0237371 |
| $10^{-2}$ | $2\mu^2$ | $10^{-4}$ | 0.960 | 47.040 | 0.0004717 | 0.0300839 |
| $10^{-3}$ | $\mu^2/2$ | $5 \times 10^{-7}$ | 1.994 | 995.006 | 0.0001772 | 0.0023721 |
| $10^{-3}$ | $2\mu^2$ | $10^{-6}$ | 0.996 | 497.004 | 0.0000509 | 0.0029970 |
| $10^{-4}$ | $\mu^2/2$ | $5 \times 10^{-9}$ | 1.999 | 9995.001 | 0.0000178 | 0.0002372 |
| $10^{-4}$ | $2\mu^2$ | $10^{-8}$ | 1.000 | 4997.000 | 0.0000051 | 0.0002996 |
| $10^{-5}$ | $\mu^2/2$ | $5 \times 10^{-11}$ | 2.000 | 99995.000 | 0.0000018 | 0.0000237 |
| $10^{-5}$ | $2\mu^2$ | $10^{-10}$ | 1.000 | 49997.000 | 0.0000005 | 0.0000300 |

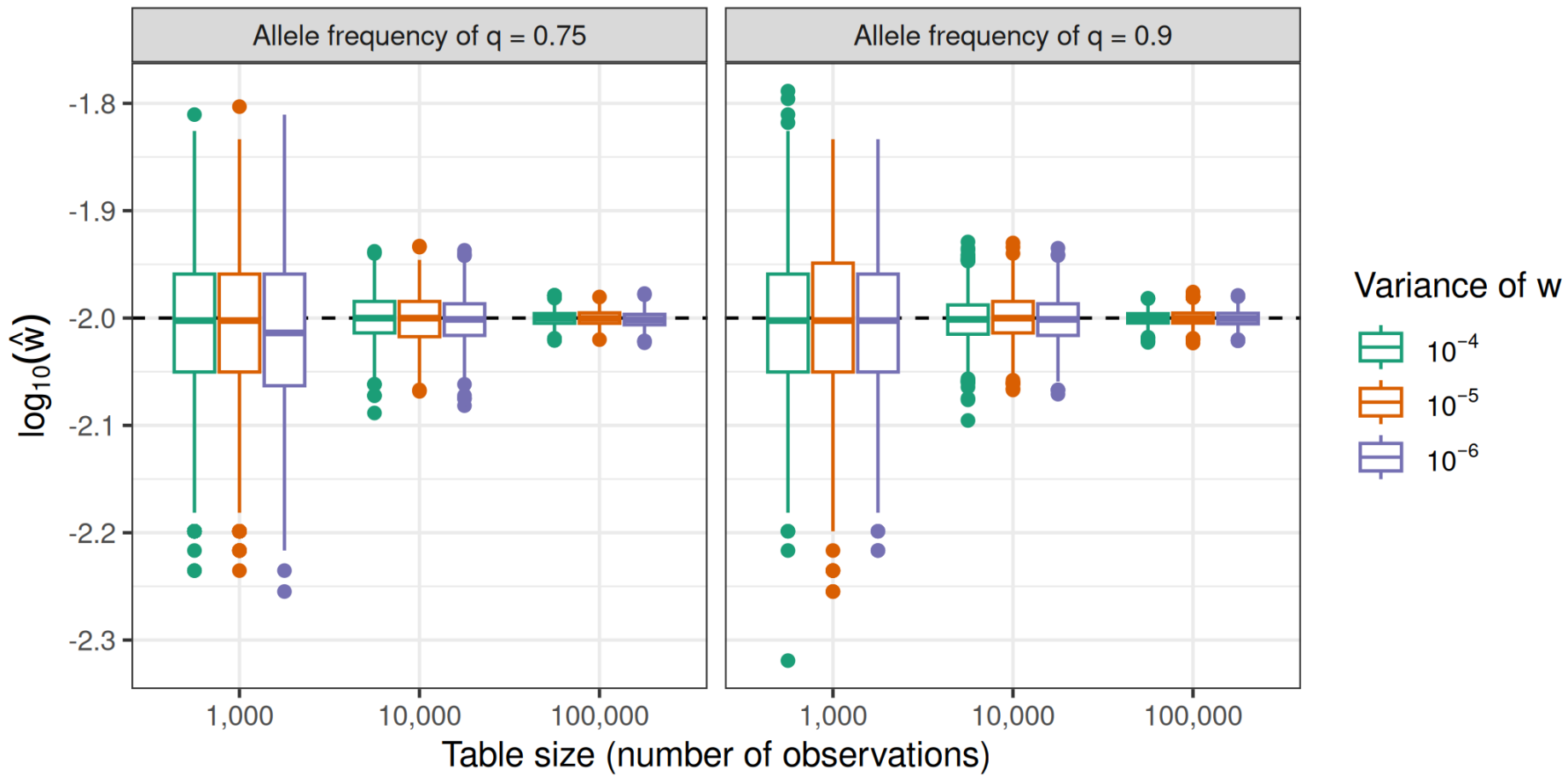

Figure 3: Sensitivity of overdispersion when estimating $w$. Refer to Figure 1 for the Beta distributions used. Each boxplot is based on 1,000 simulations (refer to text). The dashed line shows the mean of $w$, i.e., 0.01.

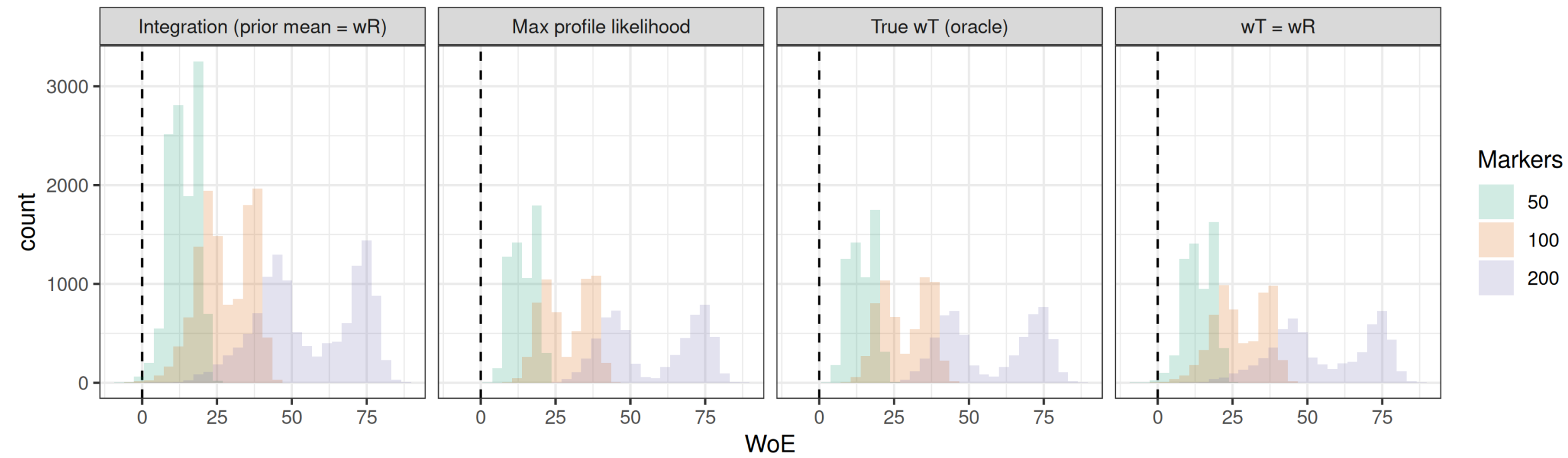

Figure 4: Cases where $H_1$ is true (donor to trace sample and reference sample is the same individual). Distribution of WoE values. Note that this is aggregated distributions over the different values of simulated $w_t$. The solid line is the identity line.

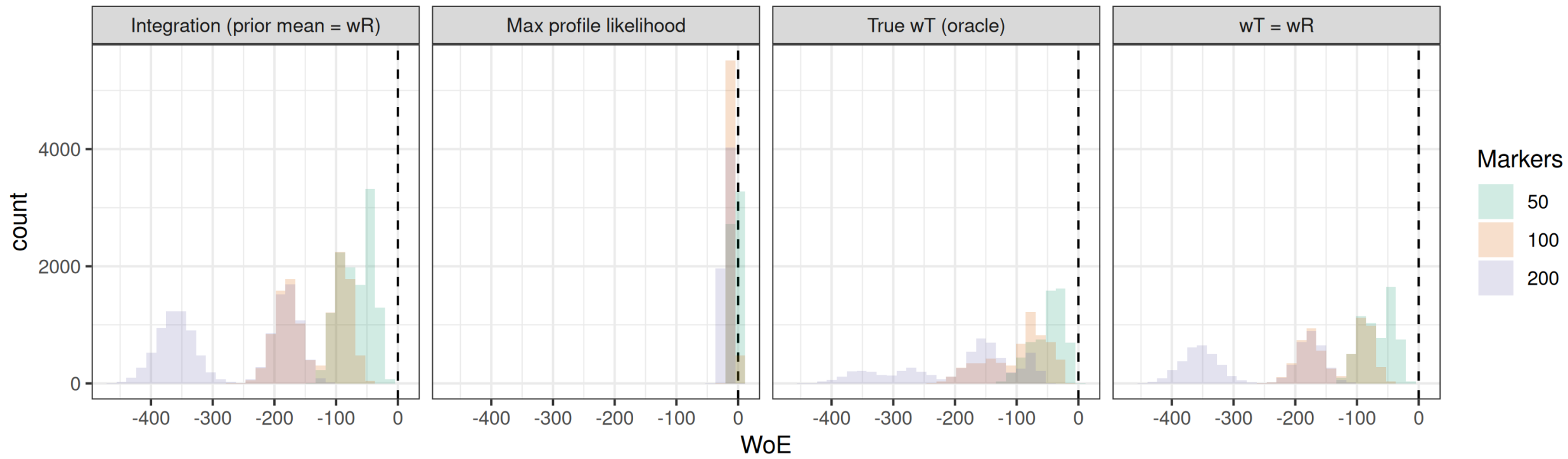

Figure 5: Cases where $H_2$ is true (donors to trace sample and reference sample are two different individuals). Distribution of WoE values. Note that this is aggregated distributions over the different values of simulated $w_t$. The solid line is the identity line.

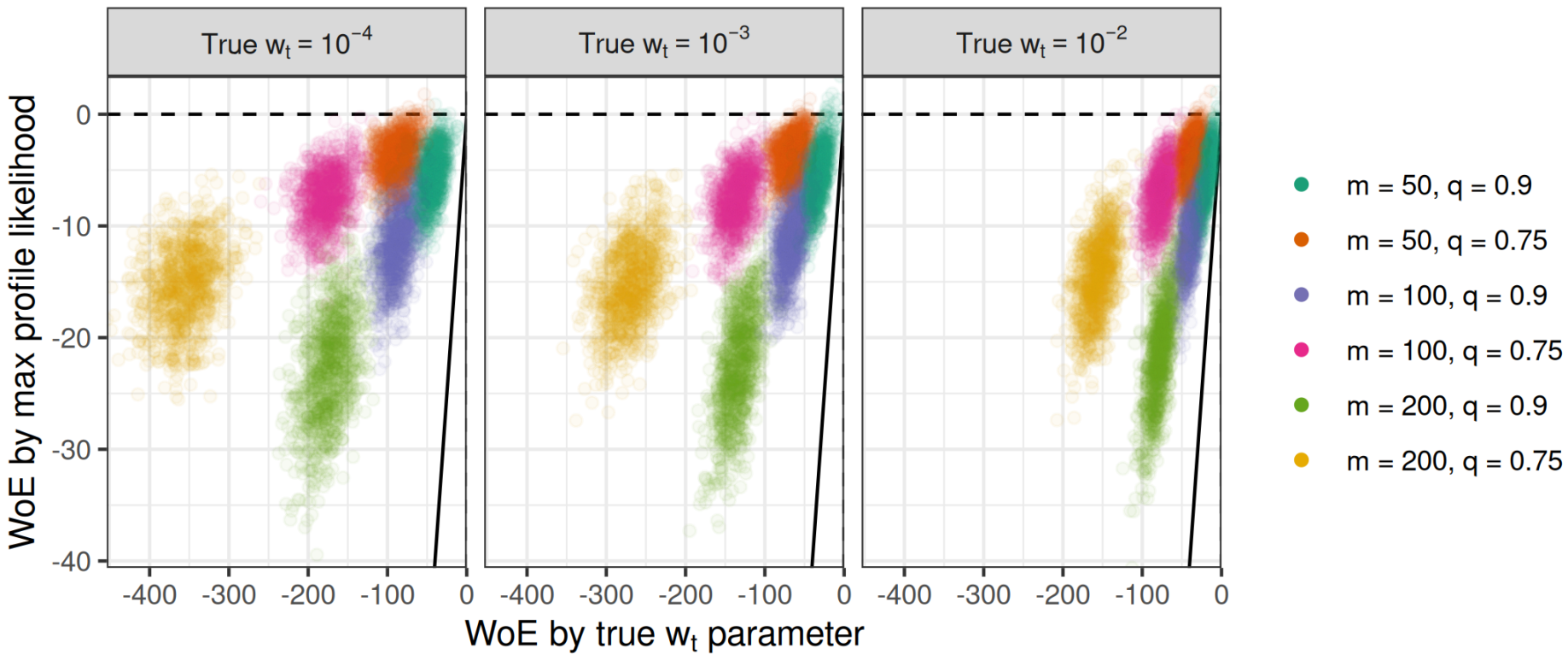

Figure 6: Cases where $H_2$ is true (donors to trace sample and reference sample are two different individuals). Results for $H_1$ true can be found in Figure S1. Comparing results if the true $w_t$ was known to those obtained using maximum of profile likehood under each hypothesis. $m$ is the number of markers and $q$ is the allele frequency. The solid line is the identity line.

Table 3: Properties of WoE for cases where $H_1$ is true (donor to trace sample and reference sample is the same individual). Refer to Table 4 for $H_2$ cases (donors to trace sample and reference sample are two different individuals). “Int” stands for integration (7) and “Max PLik” stands for maximum of profile likelihood under each hypothesis (11).

| | | | Mean WoE | | | | P(WoE > 0) [1/1,000] | | | | Min WoE | | | |
|---|---|---|---|---|---|---|---|---|---|---|---|---|---|---|
| Markers | True $w_T$ | Allele frequency | Int | Max PLik | $w_t = w_r$ | True $w_t$ | Int | Max PLik | $w_t = w_r$ | True $w_t$ | Int | Max PLik | $w_t = w_r$ | True $w_t$ |
| 50 | 1e-04 | 0.75 | 18.7 | 18.6 | 18.7 | 18.7 | 1000 | 1000 | 1000 | 1000 | 12.6 | 13.6 | 12.6 | 12.6 |
| 50 | 1e-04 | 0.90 | 11.3 | 11.3 | 11.4 | 11.4 | 1000 | 1000 | 1000 | 1000 | 3.4 | 4.8 | 3.5 | 3.5 |
| 50 | 1e-03 | 0.75 | 18.3 | 18.3 | 18.3 | 18.3 | 1000 | 1000 | 1000 | 1000 | 9.3 | 12.8 | 9.5 | 11.6 |
| 50 | 1e-03 | 0.90 | 11.1 | 11.2 | 11.1 | 11.2 | 1000 | 1000 | 1000 | 1000 | 3.8 | 5.6 | 3.8 | 4.3 |
| 50 | 1e-02 | 0.75 | 15.2 | 16.8 | 15.3 | 16.6 | 999 | 1000 | 999 | 1000 | -3.6 | 9.8 | -3.1 | 7.2 |
| 50 | 1e-02 | 0.90 | 8.3 | 10.0 | 8.4 | 9.9 | 972 | 1000 | 979 | 1000 | -7.7 | 2.1 | -7.2 | 0.5 |
| 100 | 1e-04 | 0.75 | 37.4 | 37.3 | 37.4 | 37.4 | 1000 | 1000 | 1000 | 1000 | 27.9 | 29.1 | 28.0 | 28.0 |
| 100 | 1e-04 | 0.90 | 22.7 | 22.6 | 22.7 | 22.7 | 1000 | 1000 | 1000 | 1000 | 12.5 | 13.7 | 12.7 | 12.7 |
| 100 | 1e-03 | 0.75 | 36.9 | 37.0 | 36.9 | 37.0 | 1000 | 1000 | 1000 | 1000 | 24.9 | 28.9 | 25.1 | 27.3 |
| 100 | 1e-03 | 0.90 | 22.2 | 22.3 | 22.2 | 22.3 | 1000 | 1000 | 1000 | 1000 | 10.6 | 13.7 | 11.0 | 12.8 |
| 100 | 1e-02 | 0.75 | 30.7 | 33.6 | 30.8 | 33.5 | 1000 | 1000 | 1000 | 1000 | 11.2 | 24.4 | 11.9 | 22.8 |
| 100 | 1e-02 | 0.90 | 16.3 | 19.6 | 16.6 | 19.5 | 993 | 1000 | 995 | 1000 | -3.3 | 8.6 | -2.1 | 8.3 |
| 200 | 1e-04 | 0.75 | 75.0 | 74.9 | 75.0 | 75.0 | 1000 | 1000 | 1000 | 1000 | 62.8 | 64.2 | 62.8 | 62.8 |
| 200 | 1e-04 | 0.90 | 45.4 | 45.4 | 45.4 | 45.4 | 1000 | 1000 | 1000 | 1000 | 30.7 | 30.7 | 30.7 | 30.7 |
| 200 | 1e-03 | 0.75 | 73.7 | 74.0 | 73.7 | 73.9 | 1000 | 1000 | 1000 | 1000 | 60.4 | 63.1 | 60.6 | 62.5 |
| 200 | 1e-03 | 0.90 | 44.3 | 44.7 | 44.4 | 44.6 | 1000 | 1000 | 1000 | 1000 | 29.6 | 31.2 | 29.9 | 31.2 |
| 200 | 1e-02 | 0.75 | 61.1 | 66.8 | 61.3 | 66.7 | 1000 | 1000 | 1000 | 1000 | 34.5 | 53.6 | 35.3 | 52.3 |
| 200 | 1e-02 | 0.90 | 33.0 | 39.1 | 33.4 | 39.0 | 1000 | 1000 | 1000 | 1000 | 2.3 | 25.0 | 3.8 | 22.8 |

Table 4: Properties of WoE for cases where $H_2$ is true (donors to trace sample and reference sample are two different individuals). Refer to Table 3 for $H_1$ cases (donor to trace sample and reference sample is the same individual). "Int" stands for integration (7) and "Max PLik" stands for maximum of profile likelihood under each hypothesis (11).

| | | | Mean WoE | | | | P(WoE $<$ 0) [1/1,000] | | | | Max WoE | | | |
|---|---|---|---|---|---|---|---|---|---|---|---|---|---|---|
| Markers | True $w_T$ | Allele frequency | Int | Max PLik | $w_t = w_r$ | True $w_t$ | Int | Max PLik | $w_t = w_r$ | True $w_t$ | Int | Max PLik | $w_t = w_r$ | True $w_t$ |
| 50 | 1e-04 | 0.75 | -89.6 | -3.7 | -88.3 | -88.3 | 1000 | 990 | 1000 | 990 | -40.2 | 1.8 | -39.8 | -39.8 |
| 50 | 1e-04 | 0.90 | -44.9 | -5.6 | -43.8 | -43.8 | 1000 | 998 | 1000 | 998 | -11.6 | 0.9 | -11.5 | -11.5 |
| 50 | 1e-03 | 0.75 | -90.4 | -3.7 | -89.1 | -67.6 | 1000 | 995 | 1000 | 995 | -42.6 | 1.4 | -42.1 | -30.4 |
| 50 | 1e-03 | 0.90 | -44.5 | -5.6 | -43.5 | -33.4 | 1000 | 989 | 1000 | 989 | -9.9 | 3.4 | -9.5 | -4.5 |
| 50 | 1e-02 | 0.75 | -90.6 | -3.6 | -89.3 | -39.0 | 1000 | 991 | 1000 | 991 | -44.6 | 2.1 | -44.0 | -13.9 |
| 50 | 1e-02 | 0.90 | -45.7 | -5.4 | -44.6 | -19.5 | 1000 | 996 | 1000 | 996 | -10.0 | 2.0 | -9.8 | -1.9 |
| 100 | 1e-04 | 0.75 | -179.6 | -7.4 | -177.0 | -177.0 | 1000 | 1000 | 1000 | 1000 | -116.7 | -0.3 | -115.5 | -115.5 |
| 100 | 1e-04 | 0.90 | -89.4 | -11.5 | -87.3 | -87.3 | 1000 | 1000 | 1000 | 1000 | -45.2 | -1.8 | -44.2 | -44.2 |
| 100 | 1e-03 | 0.75 | -180.4 | -7.6 | -177.8 | -135.0 | 1000 | 1000 | 1000 | 1000 | -120.5 | -0.4 | -119.2 | -88.1 |
| 100 | 1e-03 | 0.90 | -89.2 | -11.2 | -87.2 | -66.7 | 1000 | 1000 | 1000 | 1000 | -41.9 | -1.2 | -41.1 | -29.1 |
| 100 | 1e-02 | 0.75 | -179.5 | -7.1 | -176.8 | -77.0 | 1000 | 999 | 1000 | 999 | -120.0 | 0.1 | -118.9 | -45.6 |
| 100 | 1e-02 | 0.90 | -92.2 | -10.6 | -90.0 | -39.2 | 1000 | 1000 | 1000 | 1000 | -37.5 | -1.0 | -36.6 | -11.4 |
| 200 | 1e-04 | 0.75 | -360.0 | -15.3 | -354.8 | -354.8 | 1000 | 1000 | 1000 | 1000 | -265.4 | -5.4 | -261.5 | -261.5 |
| 200 | 1e-04 | 0.90 | -177.0 | -22.7 | -173.0 | -173.0 | 1000 | 1000 | 1000 | 1000 | -112.0 | -8.1 | -110.2 | -110.2 |
| 200 | 1e-03 | 0.75 | -359.3 | -15.2 | -354.1 | -268.9 | 1000 | 1000 | 1000 | 1000 | -258.8 | -5.4 | -255.6 | -193.1 |
| 200 | 1e-03 | 0.90 | -178.6 | -22.5 | -174.4 | -133.5 | 1000 | 1000 | 1000 | 1000 | -105.4 | -9.1 | -103.5 | -79.0 |
| 200 | 1e-02 | 0.75 | -363.4 | -14.5 | -358.0 | -156.5 | 1000 | 1000 | 1000 | 1000 | -264.3 | -3.7 | -260.7 | -101.0 |
| 200 | 1e-02 | 0.90 | -183.2 | -21.4 | -178.8 | -77.9 | 1000 | 1000 | 1000 | 1000 | -119.3 | -7.5 | -116.8 | -44.9 |

In realistic cases, the genotyping error probability of the trace sample is larger than that of the reference sample. Using too small probabilities for the genotyping error of the trace sample, by simply using that of the reference sample, $w_t = w_r$, was shown to give conservative WoEs (WoEs closer to 0) as shown in Figure S5 and Table 3 for $H_1$ cases (donor to trace sample and reference sample is the same individual) and in Figure S9 and Table 4 for $H_2$ cases (donors to trace sample and reference sample are two different individuals). Similarly, it was found that when integrating out the unknown $w_t$ (7), it gave more conservative results (WoEs closer to 0) when using too small a (mean) value of $w_t$ rather than too large a (mean) value (Supplementary Material Figure S3 and Figure S7). Restricting to priors with a mean corresponding to (the assumed known) $w_r$ (and aggregating the results from the two beta distributions with mean $w_r = 10^{-4}$ and variance $10^{-8}$ and $5 \times 10^{-9}$ cf. Table 2), gives results as presented in Figure S4 and Figure S8. From here, integration over prior distribution only considers the priors with mean equal to (the assumed known) $w_r = 10^{-4}$.

The WoE distributions from the different ways to handle uncertainty about the trace sample genotyping error probability, $w_t$, can be found in Figure 4 for simulated cases where $H_1$ is true (donor to trace sample and reference sample is the same individual) and in Figure 5 for simulated cases where $H_2$ is true (donors to trace sample and reference sample are two different individuals).
As seen, in the majority of the cases, the correct sign of the WoE was found, although the method by maximising the profile likelihood under each hypothesis (11) gives values closer to 0 for $H_2$ cases than the other methods, especially for 50 (and partly 100) independent markers. This is also highlighted when comparing the results to having oracle knowledge of the true $w_t$ as shown in Figure 6 for $H_2$ cases (and in Figure S1 for $H_1$ cases). Looking at the values of $w_t$ that maximise the profile likelihood under each hypothesis, there seem to be some concordance under $H_1$ for $H_1$ cases (Figure S2), but the same does not happen under $H_2$ for $H_2$ cases (Figure S6).

The results of calculating the empirical cross-entropy (ECE) and wrong sign of WoE are presented in Figure S10 and Figure S11, respectively. Note that none of the methods had a wrong sign of the WoE when using 200 markers (Figure S11).

As seen, maximising profile likelihood (11) performs slightly worse (Figure S10), especially for 50 and 100 markers, which is driven by some $H_2$ cases have positive WoE (Figure S11). For high values of $w_t$, the other methods' ECE values also increase (Figure S10), but that is driven by $H_1$ cases with negative WoE values.

## 4. Discussion

The sensitivity of the model against overdispersion was also investigated and it was found that it is very robust against overdispersion in estimating $w$ (Figure 3).

With this paper, the original model [5] for single-source stains on called genotype data can now handle the following situations: A) Two equal and known genotyping error probabilities (originally described in [5]); B) Two different and known genotyping error probabilities (e.g., hair trace sample and buccal swab reference sample); and C) One known and one unknown known genotyping error probability (e.g., known for buccal swab reference sample and unknown for trace sample). The `R` package `wgsLR` has also been updated making this functionality available, and the development version is available at https://github.com/mikldk/wgsLR. The method described in this paper can easily be extended to allow for both genotyping error probabilities being unknown.

It was found that an unknown genotyping error probability for the trace sample, $w_t$, could be dealt with by different methods: integrating it out (7), maximising profile likelihood (11), or simply using a plug-in estimate of $w_t = w_r$. It was found that it is more conservative to use too low values of $w_t$ rather than a too high values of $w_t$ as too high a genotyping error probability can explain genotype inconsistencies by errors rather than being due to two different individuals were the donors for the trace and reference sample (Figure S3 and Figure S7).

Normally, the trace sample can be of poor quality (or else STR DNA profiles can be obtained by PCR-CE) and the reference sample will be of good quality. Thus, in general it is expected that $w_t > w_r$. As it was shown (for fixed value $w_t = w_r$ in Figure S5 and Figure S9; for maximising profile likelihood in Figure S1 and Figure 6; and for integration in Figure S3 and Figure S7) to be more conservative to use a genotyping error probability that is too low (rather than too high), a practical recommendation could be to simply chose $w_t = w_r$, as this satisfies underestimating it under the assumption that $w_t > w_r$. This is supported by empirical cross-entropy (ECE) in Figure S10 and by WoE sign errors in Figure S11 as well as by the summaries of the WoE distributions (Table 3 and Table 4).

The method of maximising the profile likelihood under each hypothesis is simple, but it was shown that the true $w_t$ was not estimated (Figure S6 and Figure S2), and that some cases under $H_2$ (donors to trace sample and reference sample are two different individuals) had the wrong sign (Figure 6, Table 4, and Figure S11).

Thus, if many independent markers are available, a simple approach is to use the original `wgsLR` model by [5] with a single $w$, where the $w$ is chosen to be that of the reference sample (assuming that it can be determined high precision and accuracy). Alternatively, the a prior with a mean value, e.g., corresponding to $w_r$, can be used.

If duplicate samples of the trace are available, a possible extension would be to use these to estimate the trace sample genotyping error probability, $w_t$, either as a single number or in a Bayesian setting to obtain a posterior distribution that can be used in the integration instead of the prior distribution.

## References

[1] P. Gill, H. Haned, O. Bleka, O. Hansson, G. Dørum, T. Egeland, Genotyping and interpretation of STR-DNA: Low-template, mixtures and database matches – Twenty years of research and development, Forensic Science International: Genetics 18 (2015) 100–117. doi:10.1016/j.fsigen.2015.03.014.
URL http://dx.doi.org/10.1016/J.FSIGEN.2015.03.014

[2] GlobalFiler™IQC PCR Amplification Kit - User Guide, revision F (2024).
URL https://www.thermofisher.com/order/catalog/product/A43565

[3] Yfiler™Plus PCR Amplification Kit - User Guide, revision D (2024).
URL https://www.thermofisher.com/order/catalog/product/4482730

[4] O. L. Meyer, F. T. Petersen, B. T. Simonsen, J. Tfelt-Hansen, C. Børsting, Nationwide study on forensic genetic analyses in criminal cases in Denmark, Forensic Science International 377 (2025) 112629. doi:10.1016/j.forsciint.2025.112629.
URL http://dx.doi.org/10.1016/j.forsciint.2025.112629

[5] M. M. Andersen, M.-L. Kampmann, A. H. Jepsen, N. Morling, P. S. Eriksen, C. Børsting, J. D. Andersen, Shotgun DNA sequencing for human identification: Dynamic SNP selection and likelihood ratio calculations accounting for errors, Forensic Science International: Genetics 74 (2025) 103146. doi:10.1016/j.fsigen.2024.103146.
URL http://dx.doi.org/10.1016/j.fsigen.2024.103146

[6] M.-L. Kampmann, C. Børsting, A. H. Jepsen, M. M. Andersen, C. I. Aagreen, B. Poggiali, C. G. Jønck, N. Morling, J. D. Andersen, Preparing for shotgun sequencing in forensic genetics - Evaluation of DNA extraction and library building methods, Forensic Science International: Genetics 76 (2025) 103234. doi:10.1016/j.fsigen.2025.103234.
URL http://dx.doi.org/10.1016/j.fsigen.2025.103234

[7] P. Mostad, A. Tillmar, D. Kling, Improved computations for relationship inference using low-coverage sequencing data, BMC Bioinformatics 24 (1) (Mar. 2023). doi:10.1186/s12859-023-05217-z.
URL http://dx.doi.org/10.1186/s12859-023-05217-z

[8] R. Nguyen, J. D. Kapp, S. Sacco, S. P. Myers, R. E. Green, A computational approach for positive genetic identification and relatedness detection from low-coverage shotgun sequencing data, Journal of Heredity 114 (5) (2023) 504–512. doi:10.1093/jhered/esad041.
URL http://dx.doi.org/10.1093/jhered/esad041

[9] F. Ouerghi, D. E. Krane, M. D. Edge, On forensic likelihood ratios from low-coverage sequencing, Forensic Science International: Genetics 79 (2025) 103302. doi:10.1016/j.fsigen.2025.103302.
URL http://dx.doi.org/10.1016/j.fsigen.2025.103302

[10] R Core Team, R: A Language and Environment for Statistical Computing, R Foundation for Statistical Computing, Vienna, Austria (2024).
URL https://www.R-project.org/

[11] H. Wickham, M. Averick, J. Bryan, W. Chang, L. D. McGowan, R. François, G. Grolemund, A. Hayes, L. Henry, J. Hester, M. Kuhn, T. L. Pedersen, E. Miller, S. M. Bache, K. Müller, J. Ooms, D. Robinson, D. P. Seidel, V. Spinu, K. Takahashi, D. Vaughan, C. Wilke, K. Woo, H. Yutani, Welcome to the tidyverse, Journal of Open Source Software 4 (43) (2019) 1686. doi:10.21105/joss.01686.

## 5. Supplementary material

### *5.1. Unknown genotyping error probabilities*

#### *5.1.1. $H_1$ cases (donor to trace sample and reference sample is the same individual)*

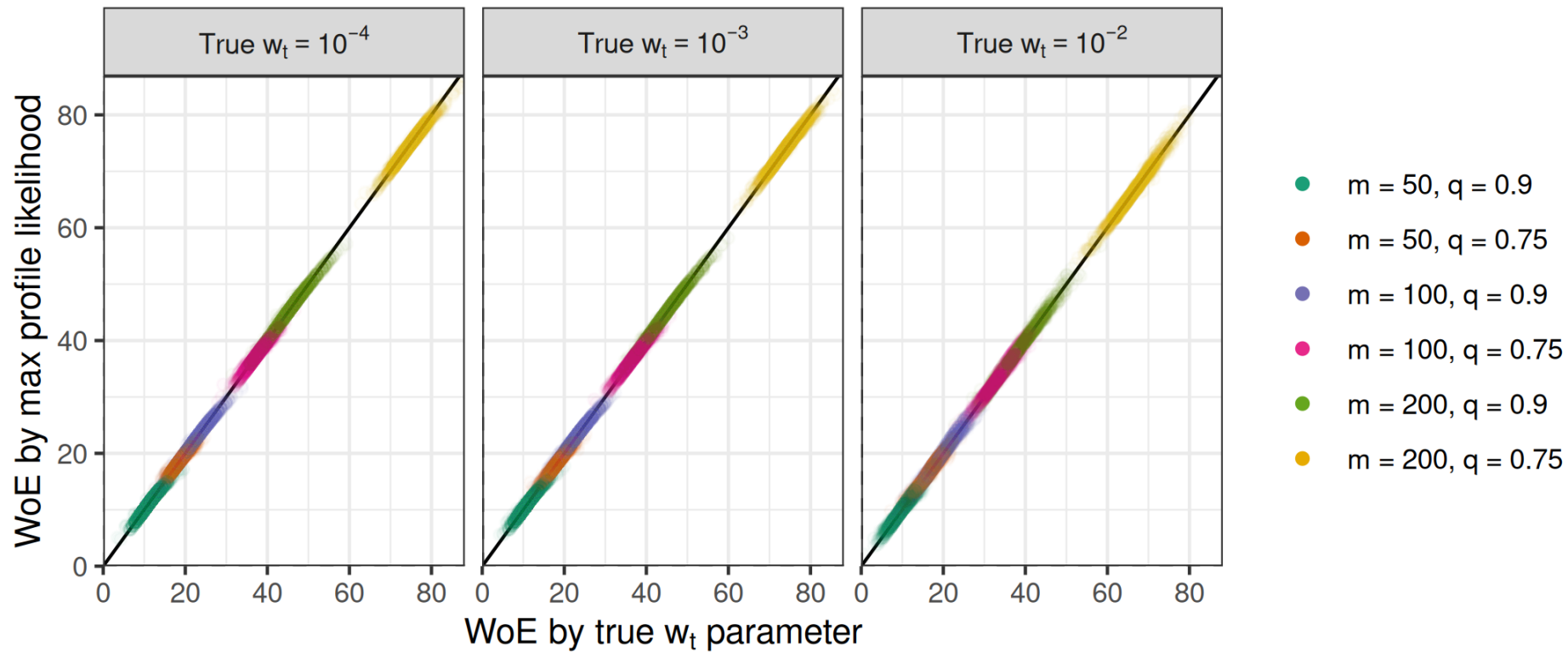

Figure S1: Cases where $H_1$ is true (donor to trace sample and reference sample is the same individual). Results for $H_2$ true can be found in Figure 6. Comparing results if the true $w_t$ was known to those obtained using maximum of profile likehood under each hypothesis. $m$ is the number of markers and $q$ is the allele frequency. The solid line is the identity line.

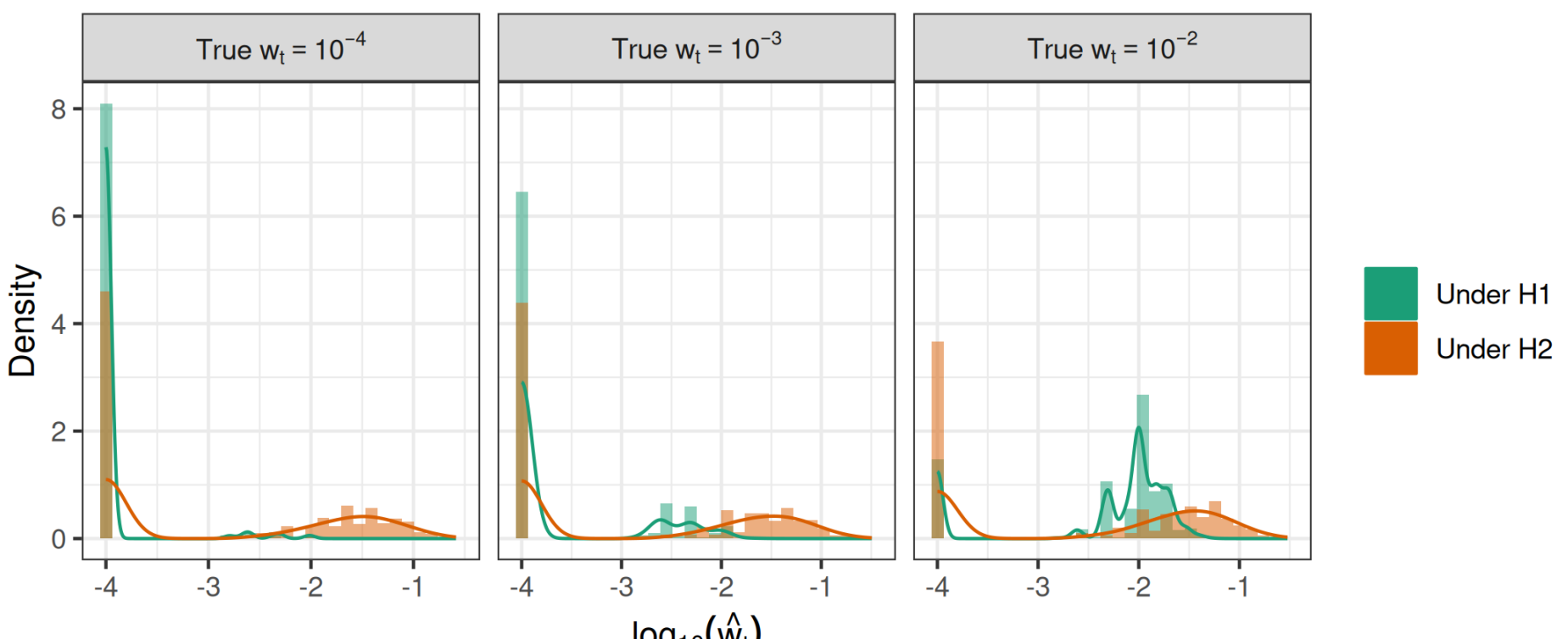

Figure S2: Cases where $H_1$ is true (donor to trace sample and reference sample is the same individual). Results for $H_2$ true can be found in Figure S6. Distribution of estimated $w_t$ values that maximise profile likehood under each hypothesis.
The solid line is the identity line.

#### *5.1.2. $H_2$ cases (donors to trace sample and reference sample are two different individuals)*

#### *5.1.3. WoE summaries*

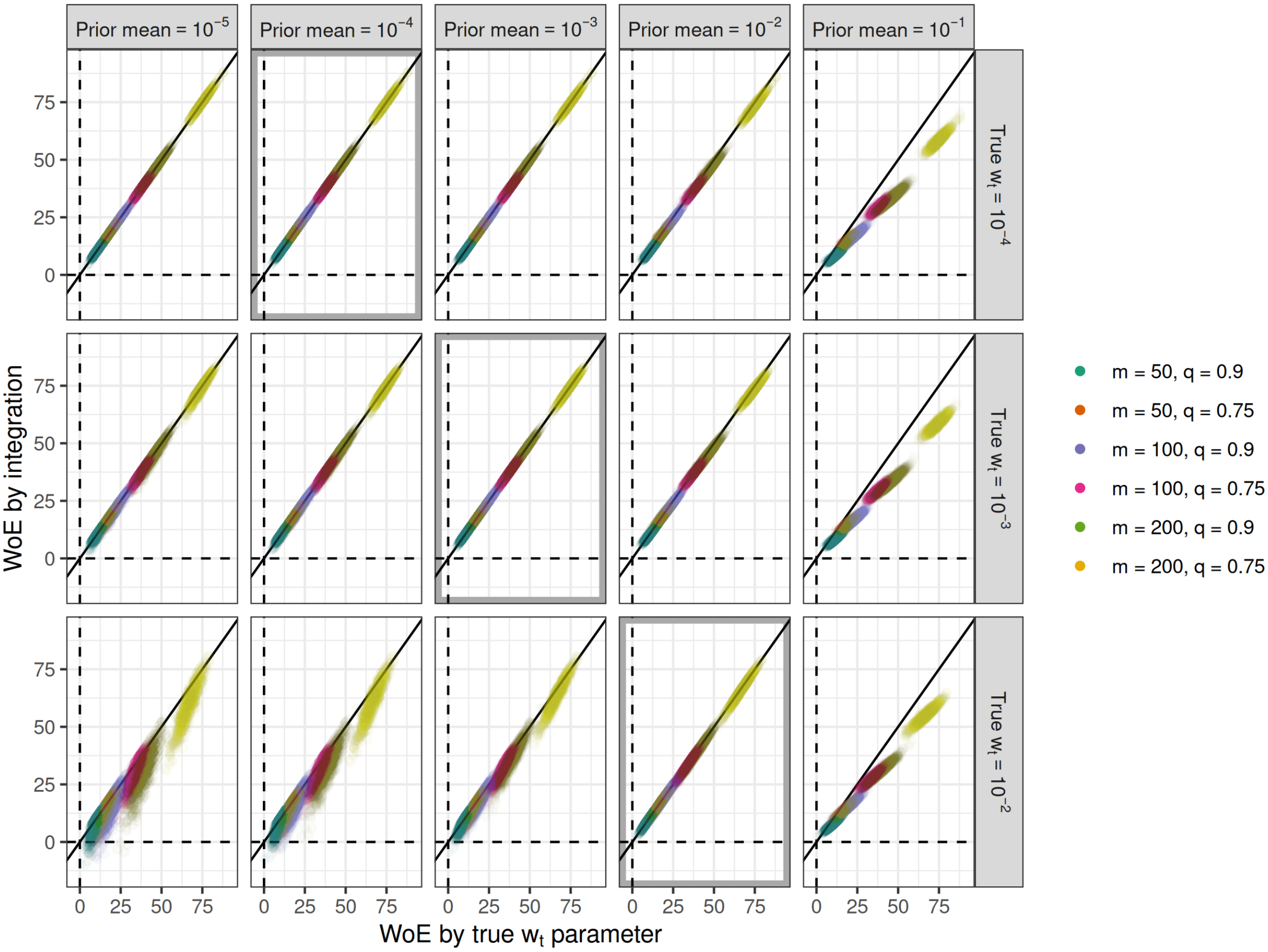

Figure S3: Cases where $H_1$ is true (donor to trace sample and reference sample is the same individual). Results for $H_2$ true can be found in Figure S7. The facets in the diagonal with grey border are those where the prior mean is the same as the true $w_t$. $m$ is the number of markers and $q$ is the allele frequency. The solid lines are the identity lines.

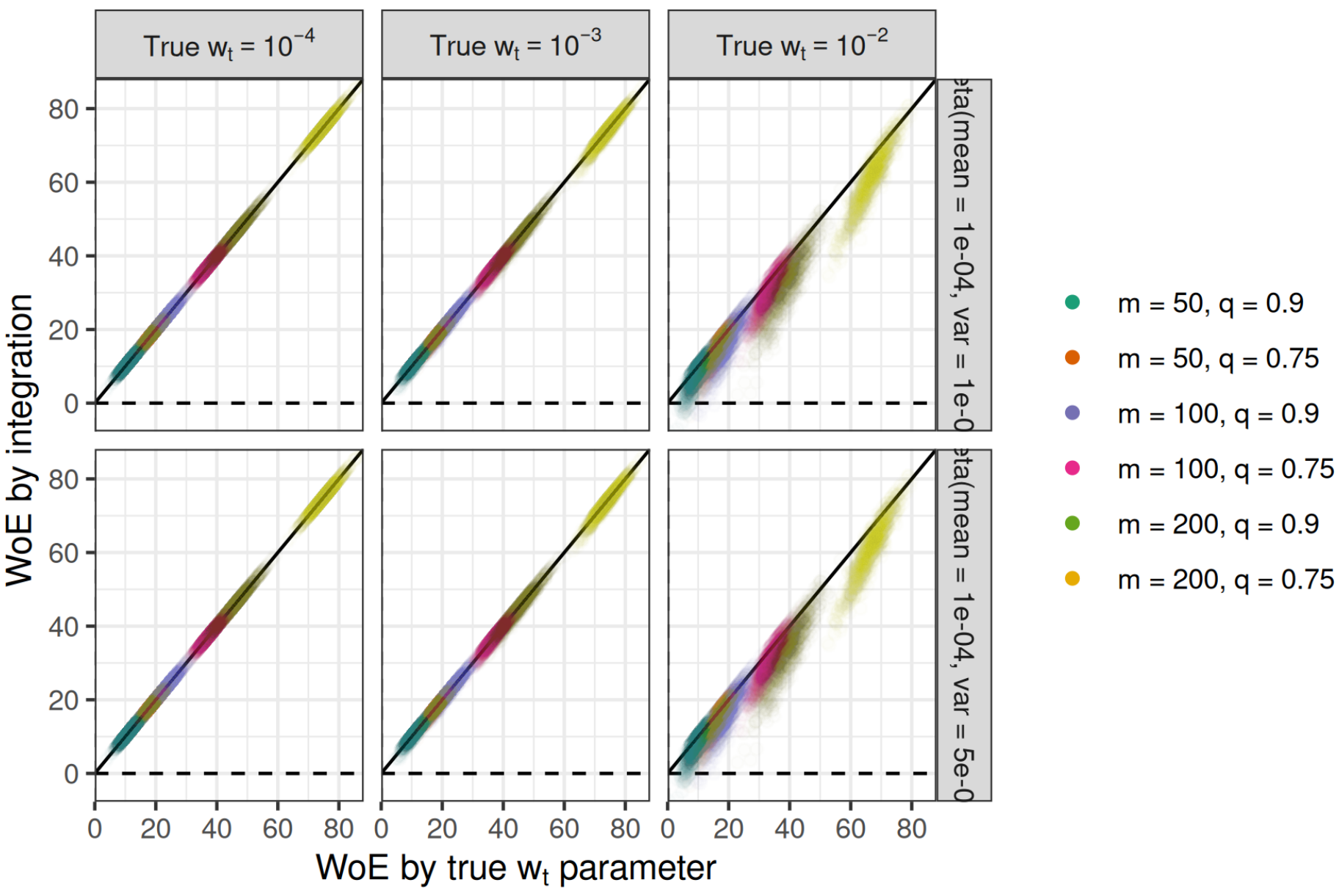

Figure S4: Cases where $H_1$ is true (donor to trace sample and reference sample is the same individual). Results for $H_2$ true can be found in Figure S8. The facets in the diagonal with grey border are those where the prior mean is the same as the true $w_t$. $m$ is the number of markers and $q$ is the allele frequency. The solid lines are the identity lines.

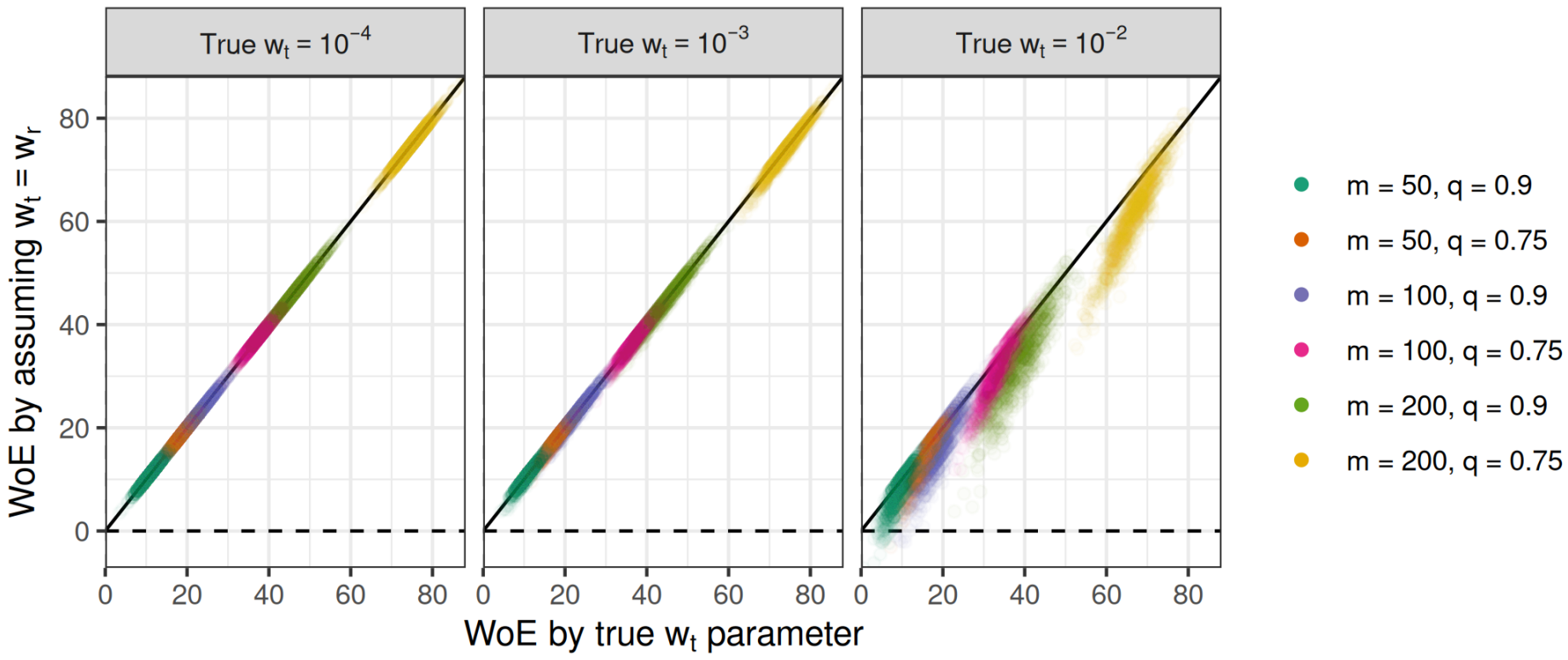

Figure S5: Cases where $H_1$ is true (donor to trace sample and reference sample is the same individual). Results for $H_2$ true can be found in Figure S9. Comparing results if the true $w_t$ was known to those obtained assuming $w_t = w_r$, i.e. using the same genotyping error probability for the trace sample as for the reference sample, i.e., using too low a value of $w_t$. $m$ is the number of markers and $q$ is the allele frequency. The solid line is the identity line.

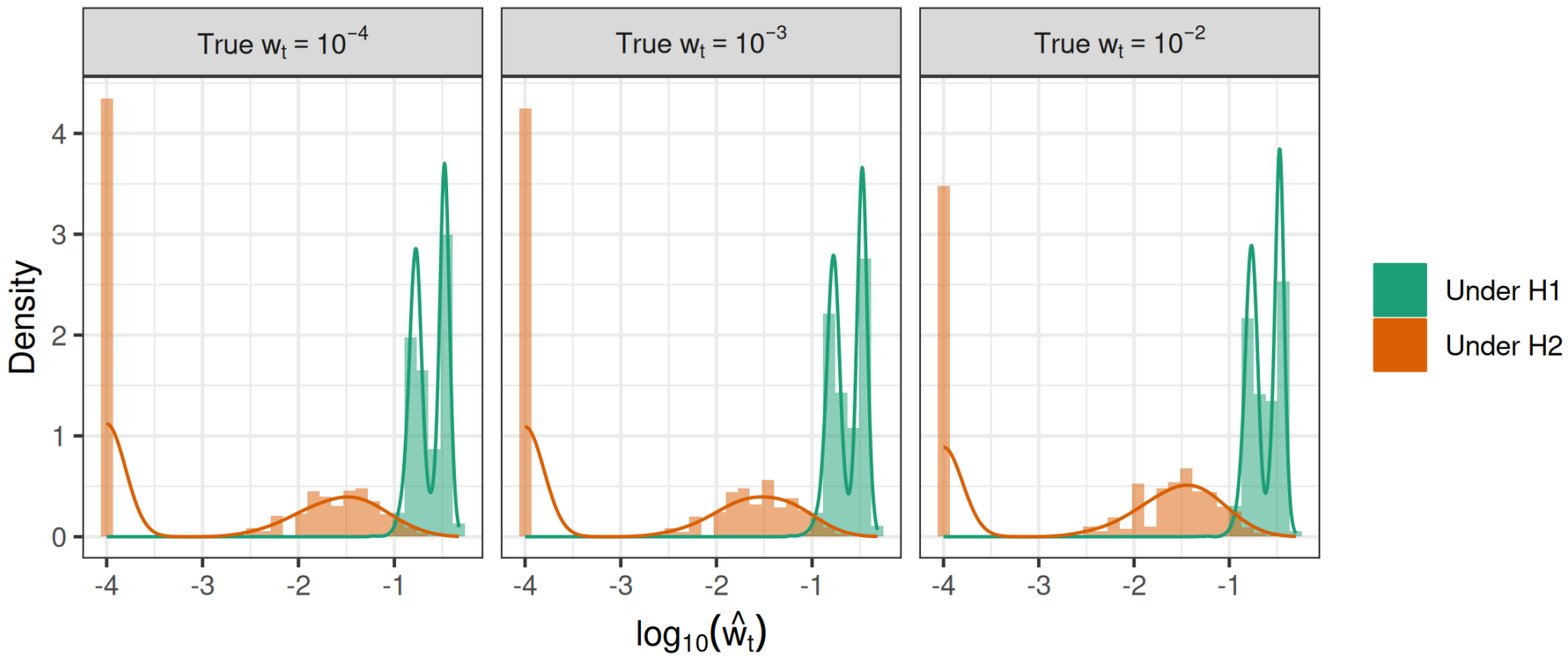

Figure S6: Cases where $H_2$ is true (donors to trace sample and reference sample are two different individuals).
Results for $H_1$ true can be found in Figure S2. Distribution of estimated $w_t$ values that maximise profile likehood under each hypothesis.
The solid line is the identity line.

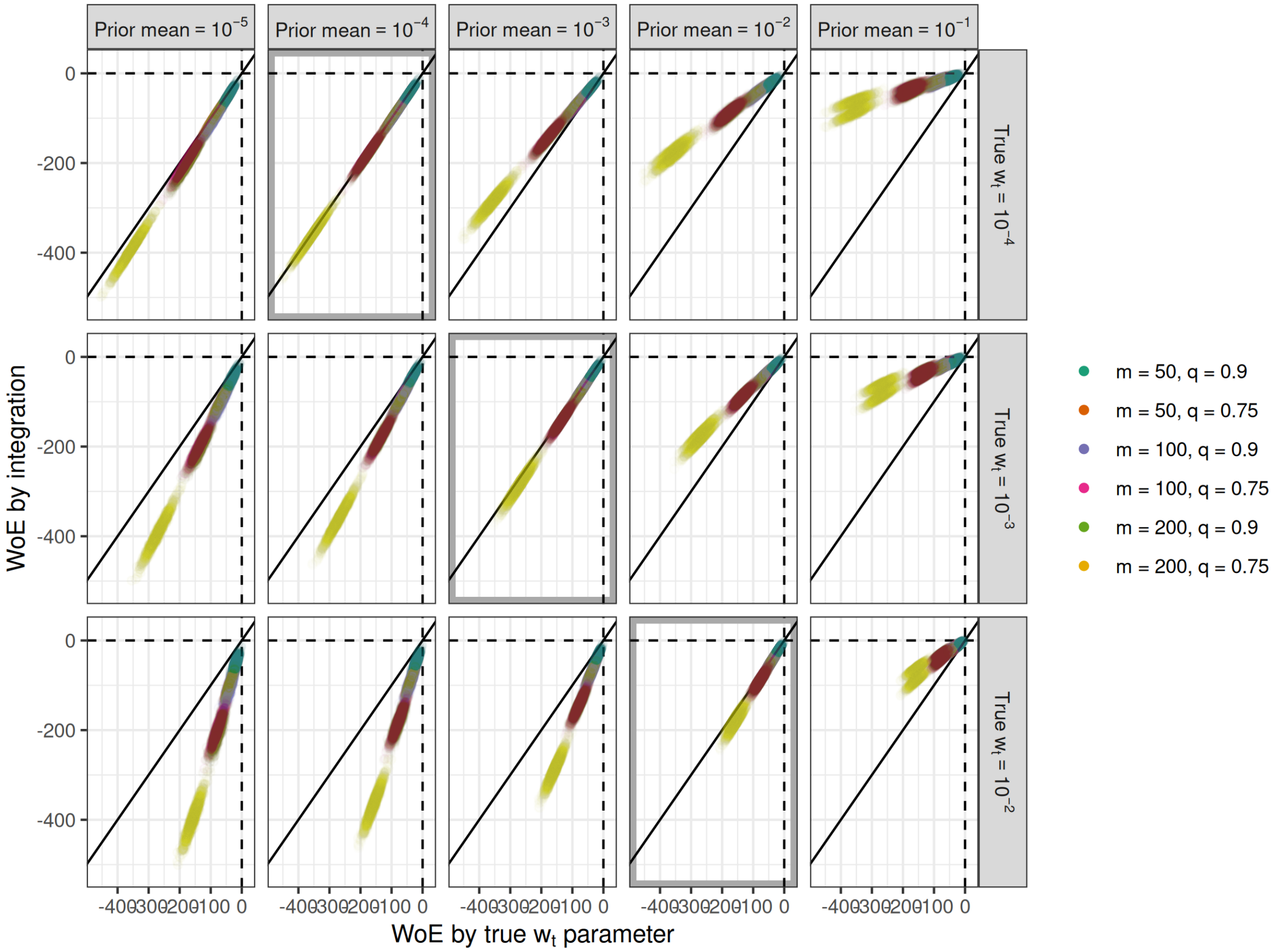

Figure S7: Cases where $H_2$ is true (donors to trace sample and reference sample are two different individuals). Results for $H_1$ true can be found in Figure S3. The facets in the diagonal with grey border are those where the prior mean is the same as the true $w_t$.
$m$ is the number of markers and $q$ is the allele frequency. The solid lines are the identity lines.

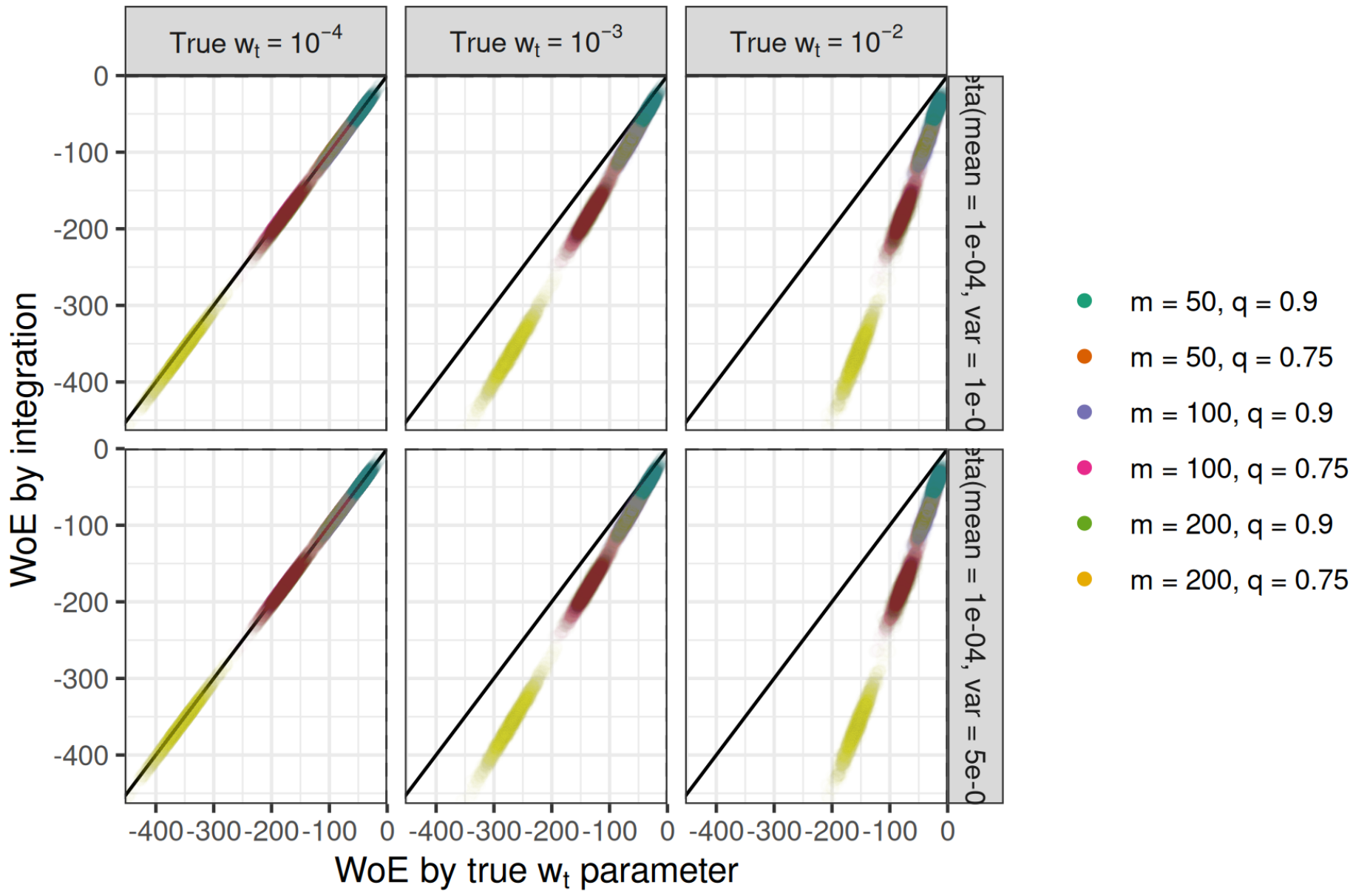

Figure S8: Cases where $H_2$ is true (donors to trace sample and reference sample are two different individuals). Results for $H_1$ true can be found in Figure S4. The facets in the diagonal with grey border are those where the prior mean is the same as the true $w_t$.
$m$ is the number of markers and $q$ is the allele frequency. The solid lines are the identity lines.

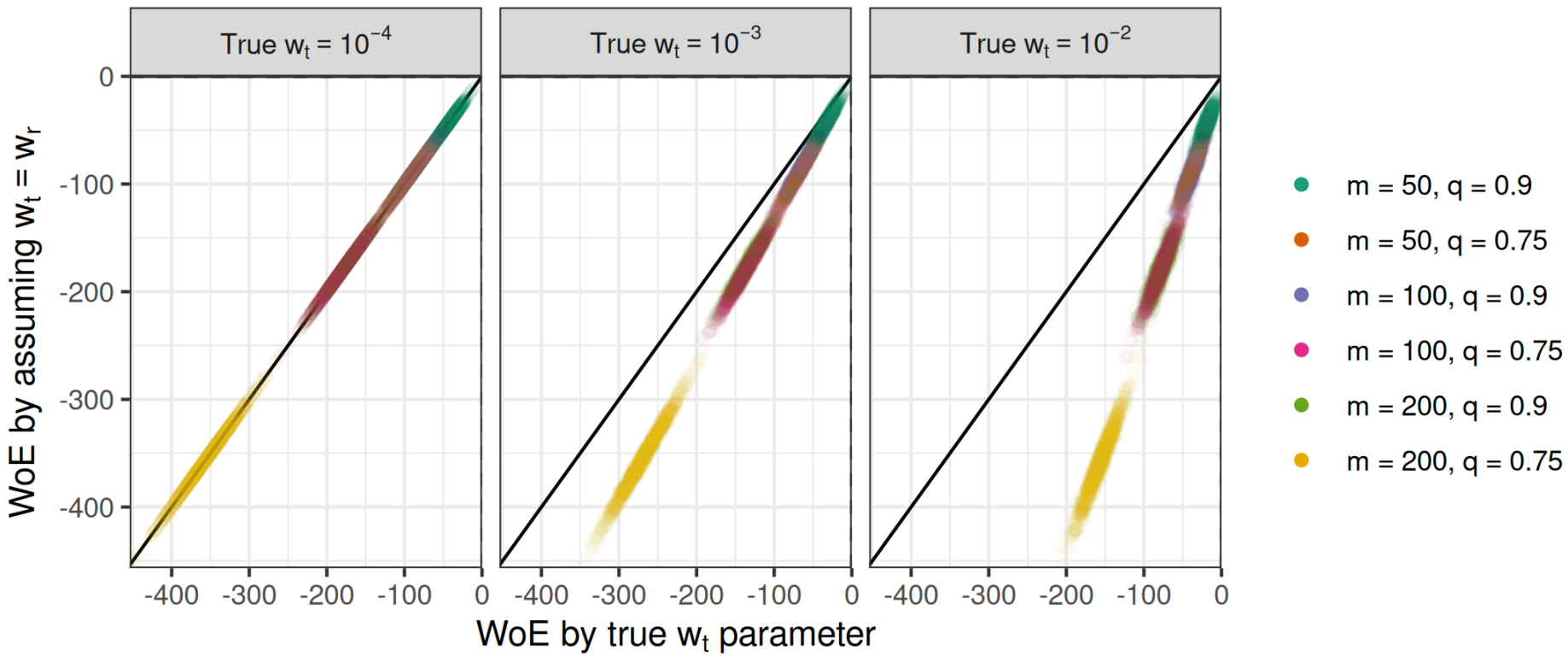

Figure S9: Cases where $H_2$ is true (donors to trace sample and reference sample are two different individuals). Results for $H_1$ true can be found in Figure S5. Comparing results if the true $w_t$ was known to those obtained assuming $w_t = w_r$, i.e. using the same genotyping error probability for the trace sample as for the reference sample, i.e., using too low a value of $w_t$. $m$ is the number of markers and $q$ is the allele frequency. The solid line is the identity line.

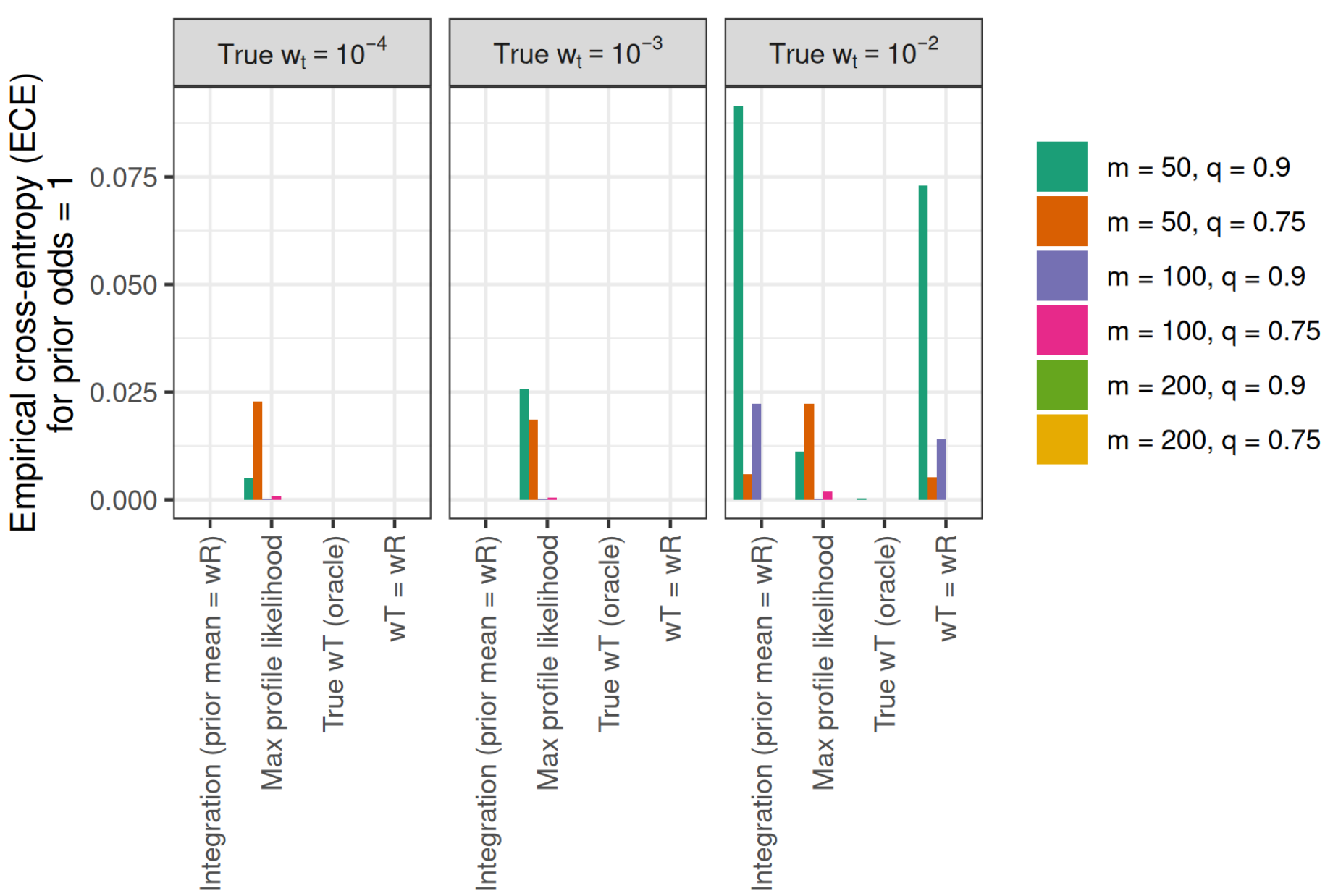

Figure S10: Comparing empirical cross-entropy (ECE) for the different methods to handle uncertainty about the trace sample genotyping error probability, $w_t$. $m$ is the number of markers and $q$ is the allele frequency.

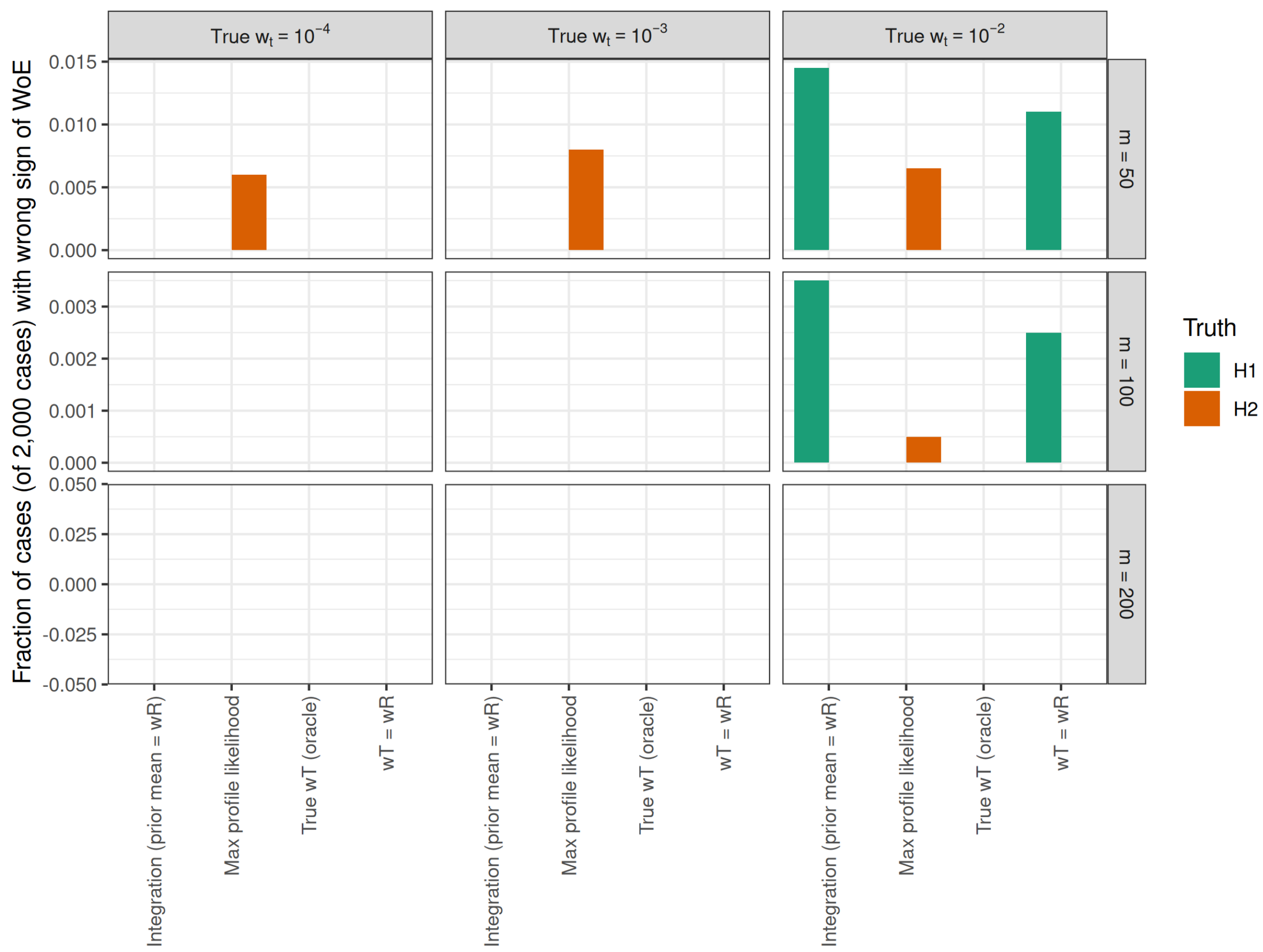

Figure S11: Comparing fraction of WoE's with wrong sign for the different methods to handle uncertainty about the trace sample genotyping error probability, $w_t$. Results are aggregated over both allele frequencies. $m$ is the number of markers.